\pdfoutput=1
\pdfmapfile{+mtpro2.map}
\documentclass[%
reprint, 
aps, prb,
superscriptaddress,
]{revtex4-2}

\usepackage{newtxtext}
\usepackage{mtpro2}

\usepackage{graphicx}
\usepackage{grffile}
\usepackage{subfigure}
\usepackage{float}

\usepackage{dcolumn}
\usepackage[super]{nth} 

\usepackage{hyperref}
\hypersetup{
    colorlinks,
    linkcolor={blue},
    citecolor={blue},
    urlcolor={black}
}

\usepackage{physics}
\usepackage[version=3]{mhchem}

\usepackage[dvipsnames]{xcolor}

\newcommand{\REV}[1]{#1}

\newcommand*{\ee}{{\rm e}}

\newcommand*{\dg}{{\dagger}}

\newcommand*{\vecr}{{\mathbf{r}}}

\newcommand*{\vecR}{{\mathbf{R}}}
\newcommand*{\vece}{{\mathbf{e}}}

\newcommand{\el}{{\rm{el}}}

\newcommand{\HF}{{\mathrm{HF}}}
\newcommand{\LF}{{\text{LF}}}
\newcommand{\GLF}{{\text{GLF}}}

\newcommand{\CS}{{\text{CS}}}
\newcommand{\CSHF}{{\text{CS-HF}}}

\newcommand{\LFHF}{{\text{LF-HF}}}

\newcommand{\occ}{{\text{occ}}}
\newcommand{\elec}{{\mathrm{elec}}}
\newcommand{\eff}{{\mathrm{eff}}}

\newcommand*{\abinitio}{{\textit{ab initio}} }

\usepackage[T1]{fontenc}

\begin{document}


\title{Variational Lang-Firsov approach plus M\o{}ller–Plesset perturbation theory \\
with applications to
ab initio polariton chemistry}


\author{Zhi-Hao Cui}
\email{zhcui0408@gmail.com}
\affiliation{Department of Chemistry, Columbia University, 3000 Broadway, New York, NY 10027, United States}

\author{Arkajit Mandal}
\affiliation{Department of Chemistry, Columbia University, 3000 Broadway, New York, NY 10027, United States}

\author{David R. Reichman}
\email{drr2103@columbia.edu}
\affiliation{Department of Chemistry, Columbia University, 3000 Broadway, New York, NY 10027, United States}


\begin{abstract}
We apply the Lang-Firsov (LF) transformation to electron-boson coupled Hamiltonians and variationally optimize the transformation parameters and molecular orbital coefficients to determine the ground state. M\o{}ller–Plesset  (MP-$n$, with $n = 2$ and $4$) perturbation theory is then performed on top of the optimized LF mean-field state to improve the description of electron-electron and electron-boson correlations. The method (LF-MP) is applied to several electron-boson coupled systems, including the Hubbard-Holstein model, diatomic molecule dissociation (\ce{H2}, \ce{HF}), and the modification of proton transfer reactions (malonaldehyde and aminopropenal) via the formation of polaritons in an optical cavity. We show that with a correction for the electron-electron correlation, the method gives quantitatively accurate energies comparable to exact diagonalization or coupled-cluster theory. The effect of multiple photon modes, spin polarization, and the comparison to the coherent state MP theory are also discussed. 
\end{abstract}
\maketitle
\section{Introduction}\label{sec: introduction}

Electron-boson coupled systems have been a topic of interest in condensed matter physics for many  decades~\cite{Mahan00book, Franchini2021, Basov2021Np}. The interaction between electrons and bosons, such as phonons and photons, can lead to the formation of quasiparticles, such as polarons and polaritons. A polaron is a quasiparticle consisting of an electron and its surrounding lattice deformation (phonons), whereas a polariton is a quasiparticle resulting from the strong coupling between matter excitations, such as excitons, and photons. Historically, the polaron problem has been a classic topic in condensed matter physics and plays a key role in various important phenomena, including charge transport (mobility), magnetism, and superconductivity~\cite{Mahan00book,Alexandrov10advances-polaron-book}. Recently, cavity-mediated chemical reactions have attracted renewed attention to polaritons, where experimental claims have been put forward that chemical reaction rates and pathways can be manipulated by the formation of polaritons in an optical cavity~\cite{Mandal22-review-polariton, Bhuyan23-review-polariton, Simpkins23-review-vibration, HiraiCR2023, LiARPC2022, NagarajanJACS2021, FranciscoS2021, Ruggenthaler22-polariton-ab-initio-review}, potentially providing a new dimension for fine control of chemical processes~\cite{HutchisonACID2012, FranciscoS2021, Hertzog19-review-polariton-chemistry, Bloch22-perspective-elec-photon, AhnS2023}.

Theoretically, the study of electron-boson coupled problems involves both the study of model systems as well as the development of new computational methods to treat both model and \abinitio systems. 
Several canonical lattice models have been proposed for the description of electron-phonon coupled systems~\cite{Mahan00book}, such as the Holstein, Su–Schrieffer–Heeger, and Fr\"ohlich models. For realistic materials and molecules, an \abinitio treatment is needed to predict material-specific properties. In such cases, an \abinitio electronic Hamiltonian is usually augmented with a set of harmonic phonons/photons, which are linearly coupled to the electronic degrees of freedom~\cite{Mandal22-review-polariton, Ruggenthaler22-polariton-ab-initio-review}. The boson frequencies are computed from \abinitio calculations and the coupling strength is estimated by density functional perturbation theory or taken as input parameters from experiments.
With respect to numerical method development, there has been significant progress in recent years which describes the electronic and bosonic degrees of freedom on the same footing~\cite{Mandal22-review-polariton, Ruggenthaler22-polariton-ab-initio-review, Foley23-polariton-ab-initio-review, Weight23-polariton-ab-initio-review}. These methods include density functional theory~\cite{Ruggenthaler14-QED-DFT, Schafer21-QED-DFT-functional, Tokatly13-td-qed-dft, Flick20-td-qed-dft, Yang21-td-qed-dft, Yang22-qed-tddft-gradient}, QED quantum chemistry methods (including QED coupled cluster and their relatives)~\cite{Haugland20-QED-CC, White20-eph-cc}, variational methods~\cite{Takada03-variable-displacement-LF, Wang20-NGED}, quantum Monte Carlo~\cite{Lee21-AFQMC-eph, Weight23-DMC-polariton}, density matrix renormalization group (DMRG)~\cite{Jeckelmann98-dmrg-holstein, Tezuka05-dmrg-eph}, diagrammatic and embedding-based approaches~\cite{Sandhoefer16, Reinhard19-dmet-hh-model, Pavovsevic22-embedding-QED-CC}. 

In this work, we aim to develop a numerical method that can handle both weakly and strongly coupled electron-boson coupled \abinitio systems, while maintaining an economical computational cost. To this end, we utilize the (variational) canonical transformation to handle strong coupling effects and use perturbation theory to correct the missing electron-electron and electron-boson fluctuations. Our method is based on the Lang-Firsov (LF) transformation~\cite{Lang63-Lang-Firsov-trans} for original electron-boson coupled Hamiltonians (similar ideas were used in Refs.~\cite{Silbey84-variational-2-level-model, Silbey89-review, Takada03-variable-displacement-LF, Ramakumar04-LF, Zhang15-variational-polaron, RiveraPRL2019, AshidaPRL2021, MandalJPCL2020, RiveraPRL2019, Riso22-SC-QED-HF}). The boson degrees of freedom are then averaged out and the energy is variationally optimized. After the electronic determinant is calculated, the remaining electron-electron and electron-boson interactions are recovered by a M\o{}ller-Plesset (MP) perturbative treatment. This scheme is useful for determining the ground state of the coupled system and preserves a low computational complexity.  For instance, second-order MP perturbation based on the transformed Hamiltonian has complexity of $\mathcal{O} (N_{X} N^5_{\text{orb}})$, a product of the number of boson configurations  $N_X$ and the number of orbitals $N_{\text{orb}}$, which is potentially more efficient for large molecules or materials systems compared to more expensive QED-CCSD and higher-level approaches. The method starts from the canonical transformation and Hartree-Fock and therefore does not involve approximate density functionals and can be improved systematically.

The paper is organized as follows. In Sec.~\ref{sec: method}, we introduce the model and a general \abinitio electron-boson coupled Hamiltonian, the coherent-state based Hartree-Fock and MP perturbation theory, and then introduce the variational Lang-Firsov approach and its MP perturbative corrections. In Sec.~\ref{sec: results}, we compute the ground-state properties of Hubbard-Holstein models and \abinitio molecules coupled to an optical cavity. In particular, we discussed the classification of weak and strong coupling regions of the system, the multi-boson mode effect, and the role of symmetry-breaking reference.  Finally, we summarize the main findings in Sec.~\ref{sec: conclusion}.

\section{Model and Method}\label{sec: method}

\subsection{Electron-boson coupled Hamiltonian}\label{subsec: ham}

A generic electron-boson coupled Hamiltonian can be written as,
\begin{equation}\label{eq:generic eph Ham}
\begin{split}
    \hat{H} =& \underbrace{\sum_{pq\sigma} h_{pq} a^{\dg}_{p\sigma} a_{q\sigma}}
    _{\hat{h}^{\el}} 
    + \underbrace{ \frac{1}{2}\sum_{pqrs\sigma\tau} V_{pqrs} a^{\dg}_{p\sigma} a^{\dg}_{r\tau} a_{s\tau} a_{q\sigma}}
    _{\hat{V}^{\el}} 
    \\ 
    &+ \underbrace{\sum_{x} \omega_{x} b^{\dg}_{x} b_{x}}
    _{\hat{h}^{\mathrm{p}}}
    + \underbrace{\sum_{x}\sum_{pq\sigma} g^{x}_{pq} a^{\dg}_{p\sigma} a_{q\sigma} \qty(b_{x} + b^{\dg}_{x})}
    _{\hat{h}^{\mathrm{el\text{-}p}}} 
    ,
\end{split}
\end{equation}
where the fermion operator $a^{\dg}_{p\sigma}$ creates an electron with spin $\sigma$ at orbital $p$, and the boson operator $b^{\dg}_{x}$ creates a boson (phonon or photon) at mode $x$. $h$ denotes the one-electron integral and $V$ denotes the two-electron repulsion integral; $\omega$ is the frequency of boson and $g$ is the electron-boson coupling tensor. We have assumed non-interacting bosons, which linearly couple to electrons.

In this work, we consider two special cases of Eq.~\eqref{eq:generic eph Ham}. One is the \emph{Hubbard-Holstein model}, which describes the interacting electrons in a lattice mediated by the phonon that couples to each site.
The electronic part of the model (Hubbard part) approximates the one-electron integral as a single hopping value $t$ and restricts the summation to the nearest neighbors in lattice sites,
\begin{equation}
\begin{split}
    h_{pq} &= t \delta_{q, p+\eta},\\
\end{split}
\end{equation}
where $\eta$ limits  $p$ and $q$ to be neighboring sites.
The two-electron integral is replaced by an on-site repulsion $U$, namely
\begin{equation}
\begin{split}
    V_{pqrs} = U \delta_{pq} \delta_{rs} \delta_{qr} .\\
\end{split}
\end{equation}
For the phonon-related part, every phonon shares the same frequency $\omega_x = \omega$ and an electron-phonon coupling constant $g$,
\begin{equation}\label{eq:hh model g tensor}
\begin{split}
    g^{x}_{pq} = g \delta_{xp} \delta_{pq} ,
\end{split}
\end{equation}
where the $\delta_{xp}$ means each phonon is locally attached to each lattice site.

The second Hamiltonian we discuss is an \abinitio light-matter Hamiltonian, which describes a molecular system coupled to quantized radiation inside an optical cavity.   In this work the  light-matter interaction is described within the long-wavelength approximation, as is typically done,  described by the Pauli–Fierz Hamiltonian in the dipole gauge~\cite{Ruggenthaler18-review-QED-ham, Mandal22-review-polariton, Rokaj18-self-energy-polariton, Mandal20-electron-trans-polariton},
\begin{equation}\label{eq:ab initio g tensor}
\begin{split}
    g^{x}_{pq} = -\sqrt{\frac{\omega_x}{2}} \lambda_{\text{c}}  \qty(\vece \cdot \boldsymbol{\mu})_{pq} ,
\end{split}
\end{equation}
where $\lambda_{\text{c}}$ is the coupling strength, $\vece$ is the polarization direction, and $\boldsymbol{\mu}$ is the molecular dipole integral,
\begin{equation}\label{eq:ab initio dipole}
\begin{split}
    \boldsymbol{\mu}_{pq} = \mel{\phi_{p}}{(\vecr - \sum_{I} Z_I \vecR_I)}{\phi_{q}} ,
\end{split}
\end{equation}
whose matrix element is the difference between the electronic and nuclear dipole ($Z_I$ is the nuclear charge of atom $I$ at position $\vecR_I$) in the one-particle basis $\qty{\phi}$. Besides the coupling term, often one must include the dipole self-energies for each cavity mode in the dipole gauge $\frac{1}{2}(\vece \cdot \hat{\boldsymbol{\mu}})^2$~\cite{Christian20-QED-hamiltonian, Rokaj18-self-energy-polariton}. For this, each cavity mode included in the light-matter Hamiltonian contribute~\cite{Taylor22-OL, Mandal23-polariton-multilayer}
\begin{equation}\label{eq: dipole self-energy}
\begin{split}
\frac{1}{2} \qty[\sum_{pq\sigma} \lambda_{\text{c}} \qty(\vece \cdot \boldsymbol{\mu})_{pq} a^{\dg}_{p\sigma} a_{q\sigma}] \qty[\sum_{rs\tau} \lambda_{\text{c}} \qty(\vece \cdot \boldsymbol{\mu})_{rs} a^{\dg}_{r\tau} a_{s\tau}]
\end{split}
\end{equation}
to the electronic part of the original \abinitio Hamiltonian. Then the modified one-electron integral reads,
\begin{equation}\label{eq:ab initio h1}
\begin{split}
h_{pq} = h^{\mathrm{kin}}_{pq} + v^{\mathrm{nuc}}_{pq} + \frac{1}{2} N_{\mathrm{mode}} \lambda^2_{\text{c}} \sum_{r}  \qty(\vece \cdot \boldsymbol{\mu})_{pr} \qty(\vece \cdot \boldsymbol{\mu})_{rq} .
\end{split}
\end{equation}
This includes the kinetic and nuclear-electron attractive potential, as well as the self-energy term of the cavity. Similarly, the two-electron integral is expressed as,
\begin{equation}\label{eq:ab initio h2}
\begin{split}
V_{pqrs} = V^{\mathrm{elec}}_{pqrs} + N_{\mathrm{mode}} \lambda^2_{\text{c}}  \qty(\vece \cdot \boldsymbol{\mu})_{pq} \qty(\vece \cdot \boldsymbol{\mu})_{rs}.
\end{split}
\end{equation}
\REV{We note that in Eq.~\eqref{eq: dipole self-energy}, the dipole self-energy is treated as the product of second quantized dipole operators. In literature, there is another choice that employs second quantization after the product~\cite{Foley23-polariton-ab-initio-review,Deprince21-self-energy,Mctague22-self-energy,Vu22-self-energy}. The difference between the two choices involves a difference between quadrupole and the product of two dipole integrals,}
\begin{equation}\label{eq:ab initio h1 quadrupole}
\begin{split}
\Delta h_{pq} = \frac{1}{2} N_{\mathrm{mode}} \lambda^2_{\text{c}} &\big[ \mel{\phi_p}{(\vece \cdot \vecr) (\vece \cdot \vecr)}{\phi_q}  \\
& -\sum_{r} \mel{\phi_p}{\vece \cdot \vecr}{\phi_r} \mel{\phi_r}{\vece \cdot \vecr}{\phi_q}\big].
\end{split}
\end{equation}
\REV{This difference becomes zero when the single-particle basis is complete. However, when using small basis sets and large coupling, the choice would affect the result. We primarily use Eq.~\eqref{eq:ab initio h1} in the main text, and discuss the results using Eq.~\eqref{eq:ab initio h1 quadrupole} in Appendix~\ref{app:dipole self energy}.}

\subsection{Coherent state Hartree-Fock and M\o{}ller–Plesset perturbation theory}\label{subsec: cshf csmp2}

The simplest mean-field wavefunction for electron-boson coupled systems is the product state of the electronic Slater determinant and the boson vacuum,
\begin{equation}\label{eq:product wavefunc}
    \ket{\Phi} = \ket{\Phi^{\elec}} \ket{0^{\text{p}}} .
\end{equation}
This wavefunction, however, does not account for the effect of electron-boson coupling. One way to incorporate the coupling is to introduce the origin shift of harmonic oscillators (with real coherent shifts $\qty{z_{x}}$),
\begin{equation}\label{eq:CSHF wavefunction}
    \ket{\Phi^{\text{CS}}} = \underbrace{\prod_{x} \ee^{- z_x (b_x - b^{\dg}_x) }}_{\hat{U}^{\text{CS}} } \ket{\Phi} ,
\end{equation}
where $\hat{U}^{\CS}$ is the unitary transformation for displacements of bosons,
\begin{equation}
    \hat{U}^{\text{CS} \dg} b_x \hat{U}^{\text{CS}} = b_x + z_x .
\end{equation}
This wavefunction ansatz is equivalent to applying the unitary transformation $\hat{U}^{\CS}$ to the Hamiltonian $\hat{H}$(Eq.~\eqref{eq:generic eph Ham}, spin indices are ignored for simplicity)
\begin{equation}\label{eq:H CS}
\begin{split}
    \hat{H}^{\text{CS}} =& \hat{U}^{\text{CS} \dg} \hat{H} \hat{U}^{\text{CS}} \\
    =& \hat{H} + \sum_{x} \qty[\omega_x z^2_x  + \sum_{pq} 2 z_x g^{x}_{pq} a^{\dg}_{p} a_{q} + \omega_x z_x (b_x + b^{\dg}_x)] .
\end{split}
\end{equation}
In the transformed Hamiltonian, the oscillators are at new equilibrium positions, such that the wavefunction has the simple form $\ket{\Phi}$, and gives the same expectation value as $\ket{\Phi^{\CSHF}}$,
\begin{equation}
\begin{split}
  E =&  \mel{\Phi^{\CSHF}}{\hat{H}}{\Phi^{\CSHF}} \\
  =& \mel{\Phi}{\hat{U}^{\text{CS} \dg} \hat{H} \hat{U}^{\text{CS}}}{\Phi} = \mel{\Phi}{\hat{H}^{\CS}}{\Phi} .
\end{split}
\end{equation}
Under the stationary condition $\pdv{E}{z_x} = 0$, $z_x$ is determined by the one-electron density matrix $\gamma_{qp} = \expval{a^{\dg}_p a_{q}}$,
\begin{equation}\label{eq:CSHF z expression}
    z_x = - \sum_{pq} \frac{g^{x}_{pq} \gamma_{qp}}{\omega_x} .
\end{equation}
This gives the expression for the coherent-state Hartree-Fock (CS-HF) total energy,
\begin{equation}\label{eq:CSHF total energy}
E^{\text{CS-HF}} [\gamma, z(\gamma)] = E^{\text{HF}} [\gamma] + \sum_{x} \omega_x z^2_x .
\end{equation}
Compared to the pure electronic HF theory, CS-HF modifies the electronic Fock matrix by adding the last term in Eq.~\eqref{eq:H CS},
\begin{equation}\label{eq:CSHF fock matrix}
F_{pq} [\gamma, z(\gamma)] = F^{\elec}_{pq} [\gamma] + \sum_{x} 2 z_x g^{x}_{pq} .
\end{equation}
To formulate a self-consistent field (SCF) theory, the Fock matrix is diagonalized at each iteration to determine the molecular orbital (MO) energies ($\varepsilon$) and coefficients ($C$),
\begin{equation}
\sum_{q} F_{pq} C_{qm} = C_{pm} \varepsilon_{m} .
\end{equation}
The one-electron density matrix $\gamma$ is then constructed from the  lowest $N_{\elec}$ occupied orbitals,
\begin{equation}
\gamma_{pq} = \sum^{\text{occ}}_{m} C_{pm} C^{\dg}_{mq} .
\end{equation}
The coherent shifts $z$'s are regarded as implicit variables, which are determined via Eq.~\eqref{eq:CSHF z expression} at each iteration. The new $\gamma$ and $z$ are used to construct the Fock matrix through Eq.~\eqref{eq:CSHF fock matrix} in the next iteration.  At convergence, the density matrix $\gamma$ is unchanged and the corresponding stationary condition of $z$ is always satisfied.  \REV{The CS-HF is equivalent to the QED-HF when applying to the Pauli-Fierz Hamiltonian, although QED-HF is typically written in the coherent state basis and the resulting Hamiltonian is subtracted with the expectation value of the dipole operator~\cite{Haugland20-QED-CC}. }

Based on the CS-HF treatment, we can apply M\o{}ller-Plesset (MP) perturbation theory to obtain the correlation energy. Here, we briefly review the results from Refs.~\cite{Lee21-AFQMC-eph, White20-eph-cc}.
The \nth{0} order Hamiltonian is chosen as the Fock operator (Eq.~\eqref{eq:CSHF fock matrix}) 
with additional quadratic boson terms,
\begin{equation}
\hat{H}^{(0)} = \sum_{pq} F_{pq} a^{\dg}_p a_q 
+ \sum_{x} \omega_x z^2_x + \sum_{x} \omega_x b^{\dg}_x b_x.
\end{equation}
The corresponding \nth{0} order wavefunction is $\ket{\Phi}$,
\begin{equation}
\hat{H}^{(0)} \ket{\Phi} = E^{(0)} \ket{\Phi} ,
\end{equation}
with  \nth{0} order energy 
$E^{(0)} = \sum^{\text{occ}}_{m} \varepsilon_m + \sum_{x} \omega_x z^2_x$.
Note that $E^{(0)}$ is different from the CS-HF energy in Eq.~\eqref{eq:CSHF total energy}. 
The remaining part (namely, the fluctuation potential) is then,
\begin{equation}
\begin{split}
\hat{W} = \hat{H}^{\CS} - \hat{H}^{(0)} =& (\hat{V}^{\text{el}} - \hat{v}^{\HF}) + \sum_{x} \omega_x z_x (b_x + b^{\dg}_x)  \\ 
&+ \sum_{x} \sum_{pq} g^{x}_{pq} a^{\dg}_{p} a_{q} (b_x + b^{\dg}_x) .
\end{split}
\end{equation}
From Rayleigh-Schr\"odinger perturbation theory,
The first-order perturbation energy is
\begin{equation}
\begin{split}
E^{(1)} = \mel{\Phi}{\hat{W}}{\Phi} = E^{\HF} - \sum^{\occ}_{m} \varepsilon_m ,
\end{split}
\end{equation}
which recovers the CS-HF energy when added to $E^{(0)}$. For the coherent state second-order perturbation (CS-MP2), we have
\begin{equation}\label{eq:MP2 energy def}
\begin{split}
E^{(2)} = -\sum_{n} \frac{\qty|\mel{\Phi^{\text{excited}}_n}{\hat{W}}{\Phi}|^2}{E_{n} - E^{(0)}} .
\end{split}
\end{equation}
The excited determinant may contain electronic or boson excitations.  By applying the Slater-Condon rule, we have the expression
\begin{equation}
\begin{split}
E^{(2)} =& -\frac{1}{4} \sum_{ijab}  \frac{\qty|\mel{ij}{}{ab}|^2}{\varepsilon_a + \varepsilon_b - \varepsilon_i - \varepsilon_j} - \sum_{x}\sum_{i} \frac{\qty|\omega_x z_x + g^{x}_{ii}|^2}{\omega_x} \\
&- \sum_{x}\sum_{ia} \frac{\qty|g^{x}_{ia}|^2}{\omega_x +  \varepsilon_a -  \varepsilon_i } ,
\end{split}
\end{equation}
where $i, j$ are occupied MO indices, and $a, b$ are virtual MO indices. The first term includes the anti-symmetrized electron repulsion integrals in the MO basis $\mel{ij}{}{ab}$,  which comes from the electron-electron correlation (the same formula as standard MP2). The second term comes from pure boson excitations (the electronic part remains the ground-state determinant). This term, however, has no contribution if the stationary condition of $z$ (Eq.~\eqref{eq:CSHF z expression}) is satisfied. Finally, the third term arises from a single-excited determinant coupled to a boson excitation.  Higher-order excitations are zero due to the two-body nature of electron-electron interaction or the linear nature of electron-boson coupling.

\subsection{Variational Lang-Firsov approach}\label{subsec: vlf}

\begin{figure*}[!htb]
\includegraphics[width=0.99\textwidth, clip]{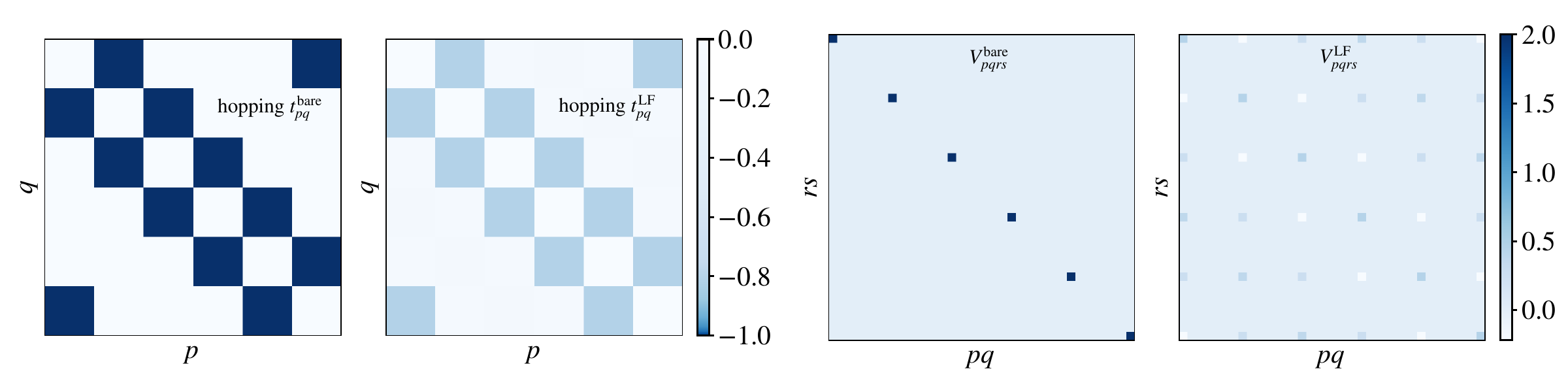}
\caption{\label{fig: lf elements}The matrix elements of hopping matrix $t_{pq}$ and
	two-electron integral $V_{pqrs}$ before and after Lang-Firsov transformation in a
	6-site 1D Hubbard-Holstein model ($U = 2$, $\omega = 0.5$, $g^2/\omega = 1.0$).
	The LF terms reduce the on-site Coulomb interaction and the (off-diagonal) hopping.}
\end{figure*}

We next use the following generalized version of the Lang-Firsov (LF) transformation,
\begin{equation}\label{eq: LF transform U}
\hat{U}^{\LF} = \exp[\sum_{xp} \lambda^{x}_{p} a^{\dg}_{p} a_{p} (b_x - b^{\dg}_x)] ,
\end{equation}
where $\lambda^{x}_{p}$ are the variational parameters. A more generic version that includes off-diagonal density coupling $\lambda^{x}_{pq}$ is possible. However, the transformed Hamiltonian does not naturally truncate, and some approximation is needed. See Appendix~\ref{app: different form of LF} for further discussion.

One can perform the LF transformation on operators using Baker–Campbell–Hausdorff (BCH) expansion,
\begin{equation}
\hat{U}^{\LF \dg} b_{x} \hat{U}^{\LF} = b_x - \sum_{p} \lambda^{x}_{p} a^{\dg}_{p} a_{p} ,
\end{equation}
\begin{equation}
\begin{split}
\hat{U}^{\LF \dg} a_{q} \hat{U}^{\LF } 
&= a_{q} \ee^{\sum_{y} \lambda^{y}_q (b_y - b^{\dg}_y)} .
\end{split}
\end{equation}
After the CS and LF transformations, the Hamiltonian becomes,

\begin{equation}
\begin{split}
\hat{H}^{\LF} =&  \hat{U}^{\CS \dg} \hat{U}^{\LF \dg} \hat{H} \hat{U}^{\LF } \hat{U}^{\CS}  \\
=& \sum_{pq} h_{pq} \ee^{\sum_y (\lambda^y_q - \lambda^y_p) (b_y - b^\dg_y)} a^{\dg}_p a_q  \\
&+ \frac{1}{2} \sum_{pqrs} V_{pqrs}\ee^{\sum_{y} (\lambda^y_q - \lambda^y_p + \lambda^y_s - \lambda^y_r) (b_y - b^\dg_y)}  a^{\dg}_p a^{\dg}_r a_s a_q   \\
& + \sum_{x} \omega_x \big(b^{\dg}_x + z_x - \sum_{p} \lambda^{x}_{p} a^{\dg}_p a_p\big) \big(b_x + z_x - \sum_{q} \lambda^{x}_{q} a^{\dg}_q a_q\big) \\
& + \sum_{xpq} g^{x}_{pq}   \ee^{\sum_y (\lambda^y_q - \lambda^y_p) (b_y - b^\dg_y) }  a^{\dg}_p a_q \\
&\, \, \, \, \, \, \, \, \times \big(b_x + b^{\dg}_x + 2 z_x - 2 \sum_{r} \lambda^x_r a^{\dg}_r a_r\big) .\\
\end{split}
\end{equation}
After averaging over the zero-boson states, the Hamiltonian becomes,
\begin{widetext}
\begin{equation}\label{eq:H LF zero boson}
\begin{split}
\hat{H}^{\LF, \elec}  =& \mel{0^{\text{p}}}{\hat{H}^{\LF}}{0^{\text{p}}} \\
= &  \sum_{x} \omega_x z^2_x  + \underbrace{\sum_{pq} \bar{h}_{pq} \ee^{-\frac{1}{2}\sum_y (\lambda^y_q - \lambda^y_p)^2} a^{\dg}_p a_q  + \sum_{xp} \qty(\omega_x \lambda^x_{p} \lambda^x_{p} - 2 \omega_x z_x \lambda^x_p) a^{\dg}_p a_p}_{\hat{h}^{\eff}_1} \\
&+ \frac{1}{2}\sum_{xpq} 2\omega_x \lambda^x_{p} \lambda^x_{q} a^{\dg}_p a^{\dg}_q a_q  a_p - \frac{1}{2} \sum_{xpqr} 4 g^{x}_{pq} \lambda^x_r \ee^{-\frac{1}{2}\sum_y (\lambda^y_q - \lambda^y_p)^2 } a^{\dg}_p  a^{\dg}_r  a_r  a_q \\
&+ \frac{1}{2} \sum_{pqrs} V_{pqrs} \ee^{-\frac{1}{2}\sum_{y} (\lambda^y_q - \lambda^y_p + \lambda^y_s - \lambda^y_r)^2} a^{\dg}_p a^{\dg}_r a_s a_q ,\\
\end{split}
\end{equation}
\end{widetext}
where
\begin{equation}
\begin{split}
\bar{h}_{pq} = h_{pq} + \sum_{x} 2 z_x g^{x}_{pq} -  \lambda^{x}_p g^{x}_{pq} - g^{x}_{pq} \lambda^{x}_q .
\end{split}
\end{equation}
During the zero-boson averaging, we have used the Franck-Condon factors for the expectation value of the boson displacement operator in the boson-number basis~\cite{Cahill69-frank-condon-factor},
\begin{equation}\label{eq:frank-condon factor}
    \mel{n}{\ee^{-z(b - b^{\dg})}}{m} =    
    \begin{cases}
      \sqrt{\frac{m!}{n!}} z^{n-m} \ee^{-\frac{z^2}{2}} L^{n-m}_m(z^2) & n\geqslant m\\
      \sqrt{\frac{n!}{m!}} (-z)^{m-n} \ee^{-\frac{z^2}{2}} L^{m-n}_n(z^2) & n < m\\
    \end{cases} 
\end{equation}
where $L^{n-m}_m$ is the associated Laguerre polynomials of degree $m$ and order $n-m$.
Physically, the transformation has two main effects: (1) It gives an exponential factor attached to one- and two-electron integrals, which makes their diagonal elements dominant; (2) It reduces the electron-electron repulsion (the penultimate term in Eq.~\eqref{eq:H LF zero boson}). In the strong coupling limit of the Hubbard-Holstein model, when $g \rightarrow \infty$, the effective $U$ becomes negative. \REV{These two effects can be seen in Fig.~\ref{fig: lf elements}, where the hopping and electron-repulsion integrals are reduced by the LF transformation.}

The ground-state energy of Eq.~\eqref{eq:H LF zero boson} can be approximated by a single HF determinant, we call this the variational LF-HF method. The corresponding Fock operator can be defined normally using the transformed integrals,
\begin{equation}
\begin{split}
\hat{F} = &\sum_{pq} \hat{h}^{\eff}_1  + \hat{v}^{\HF} .
 \end{split}
\end{equation}
The overall wavefunction is therefore parameterized by $\qty{\lambda^{x}_{p}}$, $\qty{z_x}$ and the orbital rotation between occupied and virtual blocks  $\qty{\kappa_{ia}}$,
\begin{equation}
\begin{split}
\hat{U}^{\text{rot}} = \exp(\kappa_{ia} a^{\dg}_a a_{i} - \kappa^{*}_{ia} a^{\dg}_i a_{a}) .
\end{split}
\end{equation}
Unlike the CS-HF case, the $\lambda$ parameters can not be determined analytically, and one has to variationally optimize them. The LF-HF energy reads,
\begin{equation}
\begin{split}\label{eq:LFHF total energy}
E^{\LFHF} = \sum_{x} \omega_x z^2_x + \Tr[\qty(h^{\eff} + \frac{1}{2} v^{\HF}) \gamma] .
\end{split}
\end{equation}
For computational efficiency, the analytic gradients of LF-HF energy with respect to $\lambda$, $z$, and $\kappa$ are derived in Appendix~\ref{app:analytical grad} and used during the optimization.

\REV{
We note that the diagonal form of LF parametrization is also used in the strong coupling (SC)-QED-HF~\cite{Riso22-SC-QED-HF}. 
There are subtle differences between the LF-HF and the SC-QED-HF: (1) The SC-QED-HF
is written in the diagonal basis of the dipole operator, which works well for the
case of single mode. For generic electron-boson couplings, when multiple modes exist, there is no common unitary rotation that diagonalizes all coupling tensors $g^{x}_{pq}$. In the LF-HF method, the LF  parameters are written in the basis of atomic local orbitals. (2) Although we primarily use the diagonal form of the LF transformation, the LF-HF can be generalized to the non-diagonal form (as shown in the Appendix~\ref{app: different form of LF}). (3) In LF-HF, the coherent parameters $\{z_x\}$ are explicitly written out, whereas in the SC-QED-HF the equations are written in the coherent basis.
}

\REV{As shown in Ref.~\cite{Riso22-SC-QED-HF}, after the LF parameters are optimized, not only the total energy but also the Fock matrix (MO energies) remain unchanged when the origin of the system is shifted. The same origin-invariant argument applies to the LF-HF method, since the optimized LF parameter will cancel out the shift of the dipole operator in a charged molecule.  The following perturbative treatment is based on the orbital integrals and energies, and is therefore also origin invariant.
}

\REV{Note added: During the review of our manuscript, we become aware that Li et al. have also presented a related work~\cite{Li23-VT-QED-HF} that applies variational optimized LF transformation to polariton problems.
}

\subsection{Lang-Firsov based M\o{}ller–Plesset perturbation theory}\label{subsec: lfmp2}

Similar to the CS-MP, one can formulate a Lang-Firsov M\o{}ller–Plesset (LF-MP) perturbation theory. We define the \nth{0}-order Hamiltonian as the Fock matrix in the LF-HF approach (one can obtain the Fock matrix after optimization and canonicalize it to obtain orbital energies).

The fluctuation potential is again defined as the difference between $\hat{H}^{\LF}$ and $\hat{H}^{(0)}$,
\begin{equation}
\begin{split}
\hat{W} =&  { \sum_{x} \omega_x z_x (b_x + b^{\dg}_x) }  \\
&+\sum_{pq} \bar{h}_{pq} a^{\dg}_p a_q \qty[\ee^{\sum_y (\lambda^y_q - \lambda^y_p) (b_y - b^{\dg}_y)} - \ee^{-\frac{1}{2}\sum_y (\lambda^y_q - \lambda^y_p)^2}]    \\
&- {\sum_{xp} \omega_x \lambda^x_p a^{\dg}_p a_p (b^{\dg}_x + b_x) }  \\
&+ {\sum_{xpq} g^{x}_{pq} a^{\dg}_p a_q  \ee^{\sum_y (\lambda^y_q - \lambda^y_p) (b_y - b^\dg_y) }  \qty(b_x + b^{\dg}_x) } \\
&+ \frac{1}{2}\sum_{xpq} 2\omega_x \lambda^x_{p} \lambda^x_{q} a^{\dg}_p a^{\dg}_q a_q  a_p  \\
&- \frac{1}{2} \sum_{xpqr} 4 g^{x}_{pq} \lambda^x_r a^{\dg}_p  a^{\dg}_r  a_r  a_q\ee^{\sum_y (\lambda^y_q - \lambda^y_p) (b_y - b^{\dg}_y) }   \\ 
 &+ \frac{1}{2} \sum_{pqrs} V_{pqrs} a^{\dg}_p a^{\dg}_r a_s a_q \ee^{\sum_{y} (\lambda^y_q - \lambda^y_p + \lambda^y_s - \lambda^y_r) (b_y - b^{\dg}_y)} \\
&- \hat{v}^{\HF} .
 \end{split}
\end{equation}
Within this expression, there are more terms (with exponential displacement boson operators) than the CS-MP, and therefore the number of bosons in this approach may extended to any value. In practice, one needs a truncation of the considered number of bosons $N^{\text{p}}$ (typically, 10 bosons are enough to converge the energy). The rank of electronic excitation is still fixed as there is no higher-order interactions than two-body.

Similar to the CS-MP case, the first-order correction of the LF-MP theory gives exactly the LF-HF energy. The second-order perturbation is evaluated from the definition Eq.~\eqref{eq:MP2 energy def} and utilizes the Slater-Condon rules and Frank-Condon factors. For a  collective boson configuration $X$, the phonon part of the fluctuation potential $W$ can be integrated out (using Eq.~\eqref{eq:frank-condon factor}), and the electronic part is written in the occupied ($i, j$) and virtual ($a, b$) MO orbitals,
\begin{equation}
 \begin{split}
W (0, X) =&  \bra{\Phi^{\elec}}\mel{0^{\text{p}}}{\hat{W}}{X} \ket{\Phi^{\elec}} ,\\
W^{a}_{i} (0, X) =&  \bra{\Phi^{\elec}}\mel{0^{\text{p}}}{\hat{W}}{X} \ket{\Phi^{\elec}_{ia}} ,\\
W^{ab}_{ij} (0, X) =&  \bra{\Phi^{\elec}}\mel{0^{\text{p}}}{\hat{W}}{X} \ket{\Phi^{\elec}_{ijab}} .
 \end{split}
\end{equation}
The second-order correction (LF-MP2) is then computed as,
\begin{equation}
 \begin{split}\label{eq: LF-MP2 energy}
E^{(2)} = -\sum_{X} &\left[ \frac{\qty|W (0, X)|^2}{ E_{X}} + \sum_{ia} \frac{\qty|W^{a}_{i} (0, X)|^2}{\varepsilon_a - \varepsilon_i + E_{X}} \right. \\ 
& \left.+ \sum_{ijab} \frac{\qty|W^{ab}_{ij} (0, X)|^2}{\varepsilon_a + \varepsilon_b - \varepsilon_i - \varepsilon_j + E_{X}} \right] ,
 \end{split}
\end{equation}
where the summation of boson configurations $X$ are truncated according to the predefined maximum number of bosons. For each $X$, the computational cost is $\mathcal{O} (N^5_{\text{orb}})$, the same as conventional MP2 theory. Every $X$ is independent of each other and can be embarrassingly parallelized to reduce the additional factor $N_{X}$.

Similarly, one can evaluate the third- and fourth-order MP correction in the standard manner~\cite{Cremer11-MP2-MPn-review}. For example, the fourth-order correction is
\begin{equation}
 \begin{split}
E^{(4)} =& \sum_{s>0} \sum_{t>0} \sum_{u>0} \frac{W_{0s} \bar{W}_{su} \bar{W}_{ut} {W}_{t0}}{(E_0 - E_s) (E_0 - E_t) (E_0 - E_u)} \\
&-\sum_{s>0} \sum_{t>0} \frac{W_{0s} {W}_{s0} {W}_{0t} {W}_{t0}}{(E_0 - E_s) (E_0 - E_t)^2} ,
 \end{split}
\end{equation}
where
\begin{equation}
 \begin{split}
\bar{W}_{st} = {W}_{st} - \delta_{st} W_{00} ,
 \end{split}
\end{equation}
and $s, t, u$ are compound indices of electron and boson excitations. Note that the matrix element of $W (X, Y)$ may not be Hermitian when $X$ and $Y$ have different parity.

\subsection{Computational details}\label{subsec: computational details}

The methods CS-HF, CS-MP, LF-HF, LF-MP, and exact diagonalization (ED) are implemented in the in-house program \textsc{Polar}~\cite{polar-code}. Some of the electronic routines (such as the integral evaluations and SCF procedures) are combined with \textsc{PySCF} package~\cite{Sun18pyscf,Sun20pyscf}.
Both the spin-restricted (R) and spin-unrestricted (U) versions are implemented for comparison (see below).

The equilibrium bond lengths of \ce{H2} and \ce{HF} were taken as $0.746 \text{\AA}$ and $0.918 \text{\AA}$. We used several bases (STO6G, 631G, 631**++G) for the diatomic molecules.
The optimized geometries of malonaldehyde and aminopropenal were taken from Ref.~\cite{Pavosevic22-QED-CC-proton-transfer}. The cc-pVDZ basis was used for the malonaldehyde and aminopropenal.

All calculations were converged to a change of energy smaller than $1\times 10^{-7}$ a.u.. The Lang-Firsov transformation for \abinitio systems required a set of orthogonal bases (to avoid the inclusion of atomic overlap matrix), and we choose the meta-L\"owdin orthogonalized orbitals~\cite{Sun14qmmm} in the work. We use Broyden–Fletcher–Goldfarb–Shanno (BFGS) algorithm to minimize the LF-HF energy. The initial guess of parameters is taken from CS-HF or random values.  For LF-MP, the maximum number of bosons is chosen as 16  (already converged for the systems we considered). Furthermore, the boson configurations are parallelized using MPI. 
The DMRG calculations were performed using \textsc{block2} package~\cite{Zhai2021-block2, Zhai23-block2}, which supported   customized electron-boson Hamiltonians. The bond dimension was kept as 1000 for both electron and boson sites.

\section{Numerical Results}\label{sec: results}

\subsection{Hubbard-Holstein models}

\begin{figure*}[!htb]
\subfigure[]{\includegraphics[width=0.48\textwidth, clip]{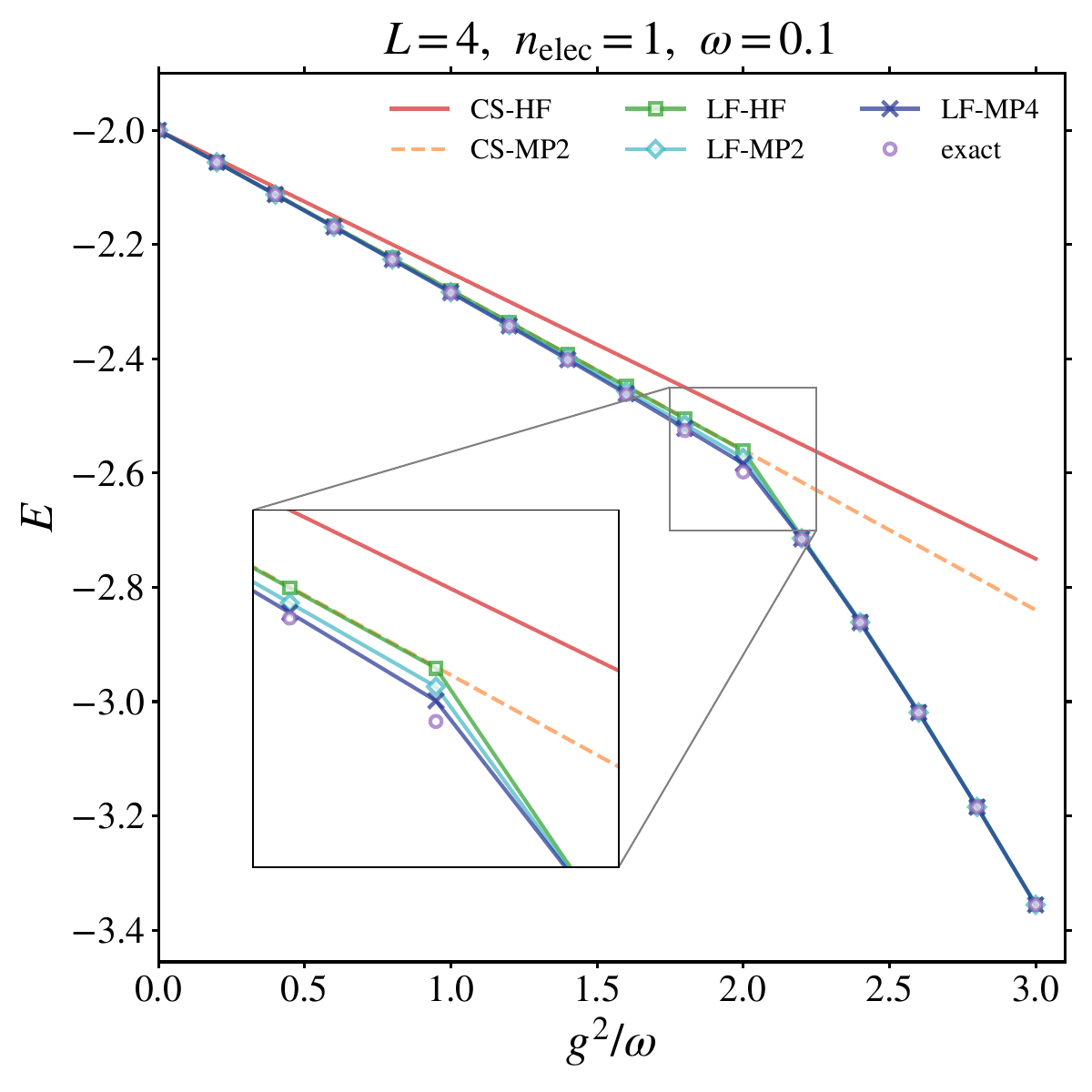}}
\subfigure[]{\includegraphics[width=0.48\textwidth, clip]{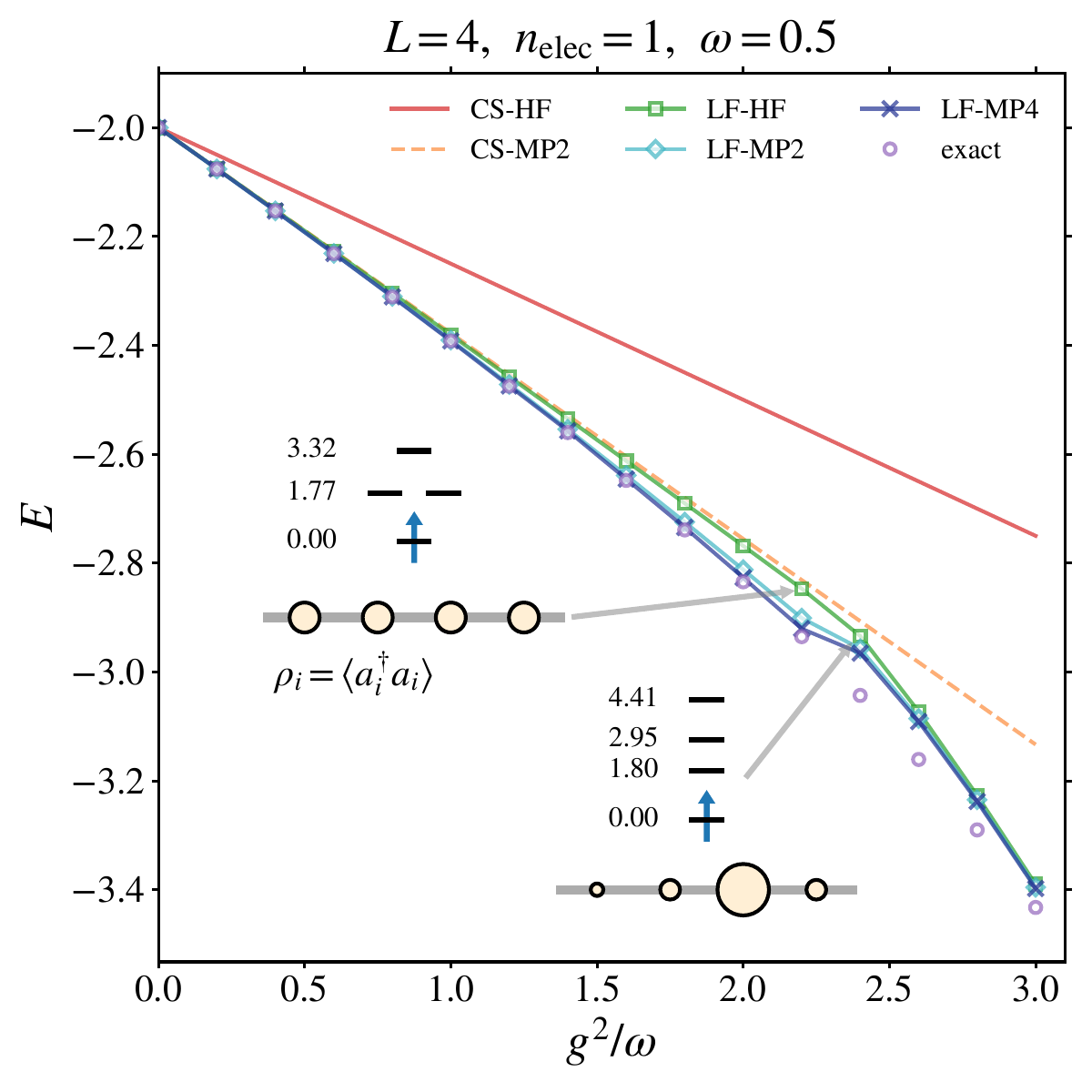}}
\subfigure[]{\includegraphics[width=0.48\textwidth, clip]{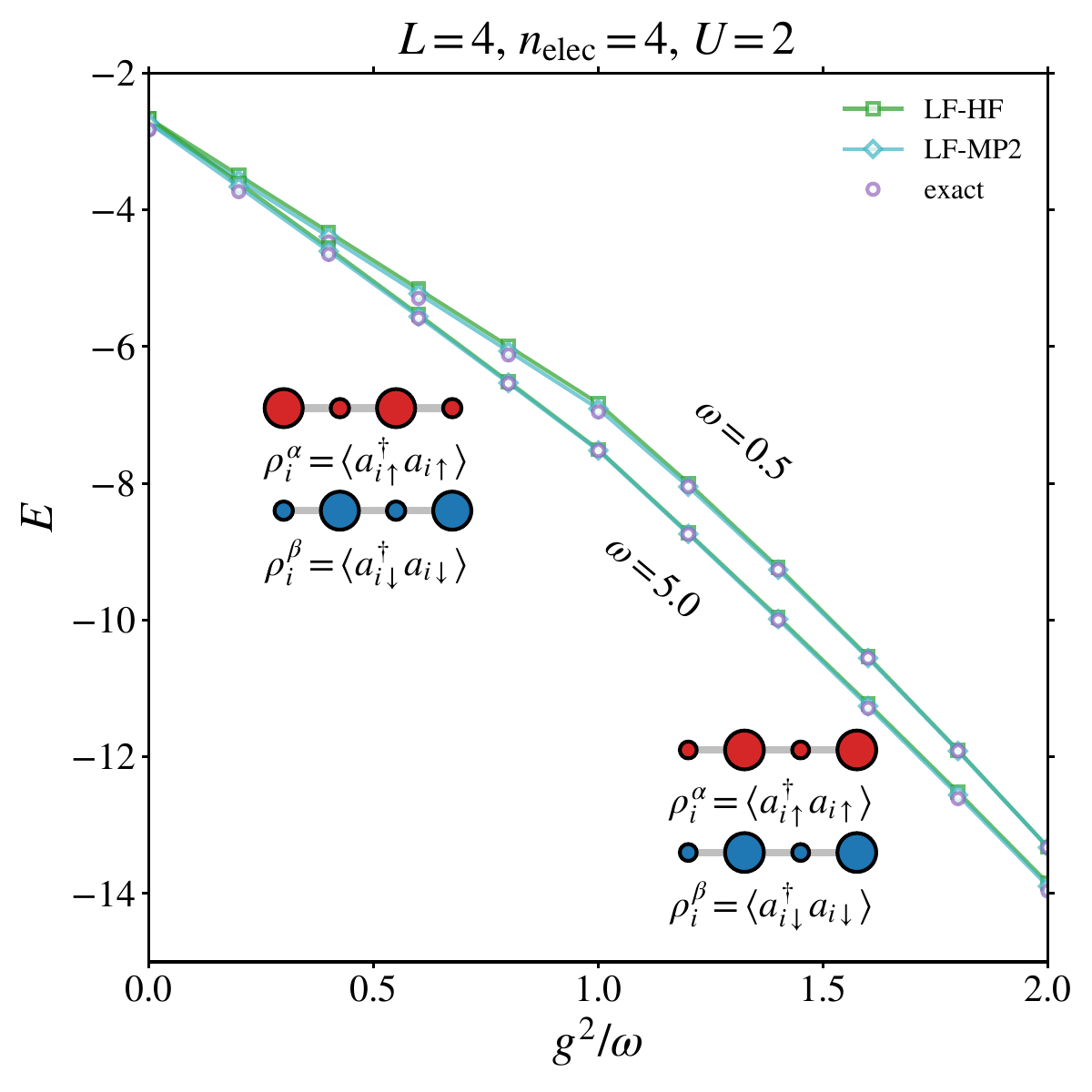}}
\subfigure[]{\includegraphics[width=0.48\textwidth, clip]{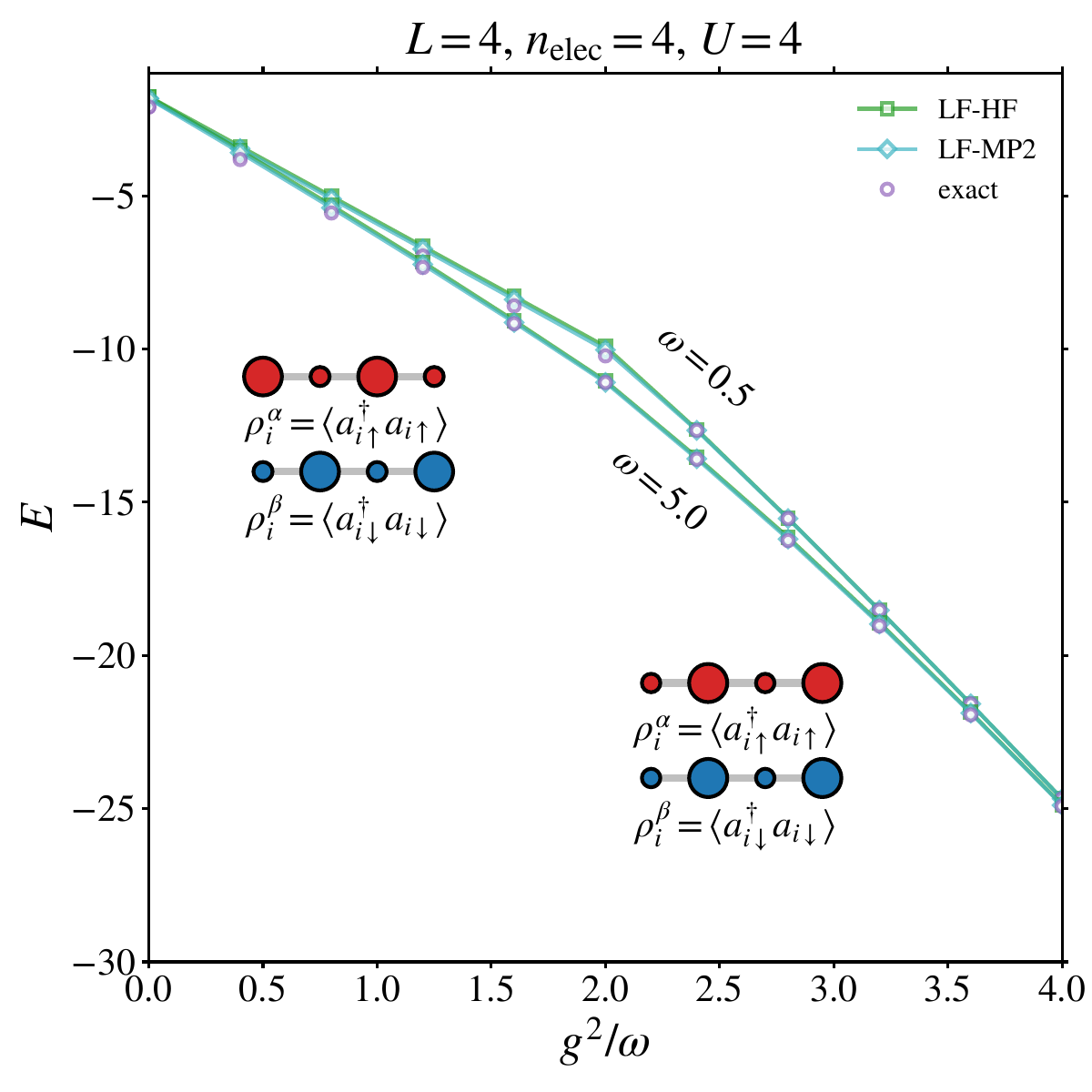}}
\caption{\label{fig: hh model}Ground-state energy of the 1D Hubbard-Holstein model, for  different coupling strengths $g^2/\omega$. The periodic lattice size $L$ is taken as 4. (a) $\omega = 0.1$, one electron. (b) $\omega = 0.5$, one electron. The electron densities $\rho_i$ at site $i$ and molecular orbital energies (relative to the lowest energy) are shown for the two coupling values (2.2 and 2.4) around the self-trapping transition. (c) 4 electrons with on-site interaction $U = 2$. (d) 4 electrons with on-site interaction $U = 4$. }
\end{figure*}
We first benchmark the ground-state energy obtained from different methods in the 1D Hubbard-Holstein model. We consider a single electron on a 4-site Hubbard-Holstein model with different phonon frequencies ($\omega$) and coupling strengths ($g^2/\omega$); the results are collected in Fig.~\ref{fig: hh model}. Note that with a single charge this reduces the Hubbard-Holstein model to the pure Holstein one.  Since the number of sites is small ($L = 4$), the ground state of the system can be found by exact diagonalization. In the adiabatic case (low frequency, $\omega = 0.1$, Fig.~\ref{fig: hh model}(a)), the LF-HF already gives a very accurate energy and naturally divides the coupling strength into two regimes: Before the slope changes ($g^2/\omega < 2$), the system is in the weak-coupling regime, whereas $g^2/\omega > 2$ defines the strong coupling regime.
The CS-HF includes the correction from the coherent shift and CS-MP2 gives the perturbative correction on top of it. These two methods, however, do not capture the transition from the weak coupling to the strong coupling. In fact, they work well in the weak-coupling limit and provide a straight line of energy when $g^2/\omega$ increases. From the inset of Fig.~\ref{fig: hh model}(a), one can see that the LF-HF gives similar accuracy to CS-MP2 in the weak-coupling regime, and captures the transition of energy in the strong-coupling one. Since the LF-HF is already very accurate, the perturbative corrections are not evident in this case. From the inset of the figure, we can see that LF-MP2 and LF-MP4 further improve the energy description and provide very similar results to exact diagonalization (maximum error is about 0.01,  0.6\% of the total energy).

We next discuss the physical interpretation of the LF approaches in Fig.~\ref{fig: hh model}(b), where $\omega = 0.5$. In this case, the weak-to-strong coupling transition happens around $g^2/\omega \sim 2.4$ and the LF-HF curve has a slope change. The error from LF-HF is larger than the model with $\omega = 0.1$ and is again similar to the error of CS-MP2 in the weak-coupling region. LF-MP2 and LF-MP4 are very accurate before the transition, but provide only small corrections after it. The error is maximized for the couplings just after the transition (largest error $\sim 2.6\%$) and becomes smaller again in the large coupling limit.
Therefore, it is worth having a deeper look at the physical origin of the slope change and the different perturbative behaviors in the two regions.
We plot the electronic density on each site $\rho_i$ of the lattice for $g^2 / \omega = 2.2$ and $2.4$. One can clearly see that the charge density becomes non-uniform after the transition. This is due to the symmetry breaking of $\lambda^{x}_{p}$ and $z_x$ during the variational optimization. This can be viewed as the ``self-trapping'' of the polaron in the strong coupling limit~\cite{Gerlach91-self-trapping-polaron-analytical}, where the variationally optimal state is localized on a few sites rather than spreading over the whole lattice.
This symmetry breaking has a direct effect on the molecular orbital energy spectrum, which is also plotted in Fig.~\ref{fig: hh model}(b). In the weak-coupling region, the orbital energy obeys the symmetry of the model Hamiltonian and has a typical ``diamond'' shaped spectrum, i.e. the second and third orbitals are degenerate. On the other hand, after the symmetry breaking, the degeneracy no longer exists and the unoccupied energy levels are shifted up to higher energies. This makes electronic excitations less accessible and reduces the correlation energy from the perturbation (because most terms involve a sizable denominator associated with electronic excitations, for example in Eq.~\eqref{eq: LF-MP2 energy}). The LF-MP2 and LF-MP4 still provide some small corrections, but cannot recover the full correlation energy as they are low-order perturbations (see also Appendix~\ref{app: different form of LF} for the correlation energies from uniform density states).
Thus, this symmetry breaking is spurious and can only be eliminated in a non-perturbative manner. 
Another perspective on this phenomenon is found via analogy to the symmetry-breaking methods in electronic structure theory.  In an approximate variational method, symmetry breaking can happen to gain a lower energy at the cost of losing good quantum numbers. This is ubiquitous in the so-called ``unrestricted'' Hartree-Fock or Kohn-Sham density functional theory. For instance, the unrestricted HF breaks the spin $S^2$ symmetry but improves the energy of the dissociation curve of diatomic molecules (see Sec.~\ref{subsec: diatom mol} for more details). Nonetheless, these symmetry-breaking approaches are often accompanied by the difficulty of including further corrections.  The exact ground state shares the same symmetry as the Hamiltonian and starting from one particular symmetry-breaking reference, it is natural that the perturbative treatment would need high orders to recover the full symmetry of the system. Alternatively, one should regard the transition region as a strongly correlated problem and start with a multi-reference state (for example, from a multi-determinant ground state, which is a linear combination of symmetry-broken states and performs perturbation on top of it). We leave this topic to future work.

We finally discuss the 1D Hubbard-Holstein model with multiple electrons in Figs~\ref{fig: hh model}(c) and (d). We consider both adiabatic ($\omega = 0.5$) and anti-adiabatic ($\omega = 5.0$) cases.  The initial guess of electron density is chosen as the antiferromagnetic (AFM) form. As shown in the figure, in the weak coupling regime, the converged spin density remains in an AFM pattern, i.e., the spin-up and spin-down densities are staggered on neighboring sites. After the transition to strong coupling, the electron-boson coupling is dominant and the converged spin densities are the same for different spin channels. Similar to the one-electron problem, the density is non-uniform in the strong coupling regime, although here the density is no longer localized at a single site. 
When boson frequency increases (anti-adiabatic case), the curve is shifted down in energy. And the error of the method with respect to exact diagonalization is also slightly smaller in the weak coupling regime. In the anti-adiabatic limit,  the boson degrees of freedom can be integrated out and are thus more easily captured by the LF-HF. 

When the on-site Coulomb interaction $U$ increases from 2 to 4, the transition point from weak to strong coupling is postponed until approximately  $g^2/\omega = 2.0$. Since the $U$ value is larger, the symmetry-breaking mean field favors the AFM state more, and the electron-boson interaction is less effective at redistributing the spin density pattern. 
Since $U = 4$ is not very large, we expect the MP2-level treatment to work well for the cases. When the electron-electron repulsion becomes extremely large, such as $U > 10$, LF-MP2 would break down in the weak coupling regime, but may still work in the strong-coupling one since the electron-boson interaction reduces the effective $U$.

\subsection{Diatomic molecules coupled to photons}\label{subsec: diatom mol}

\begin{figure*}[!htb]
\subfigure[]{\includegraphics[width=0.48\textwidth, clip]{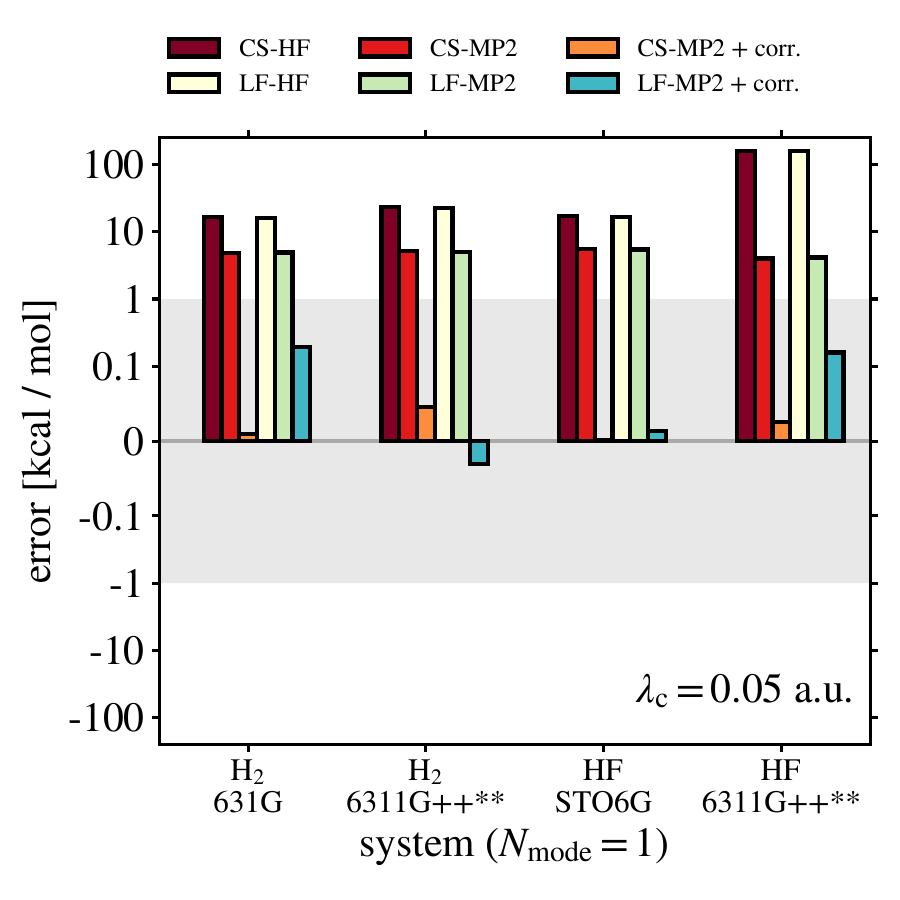}}
\subfigure[]{\includegraphics[width=0.48\textwidth, clip]{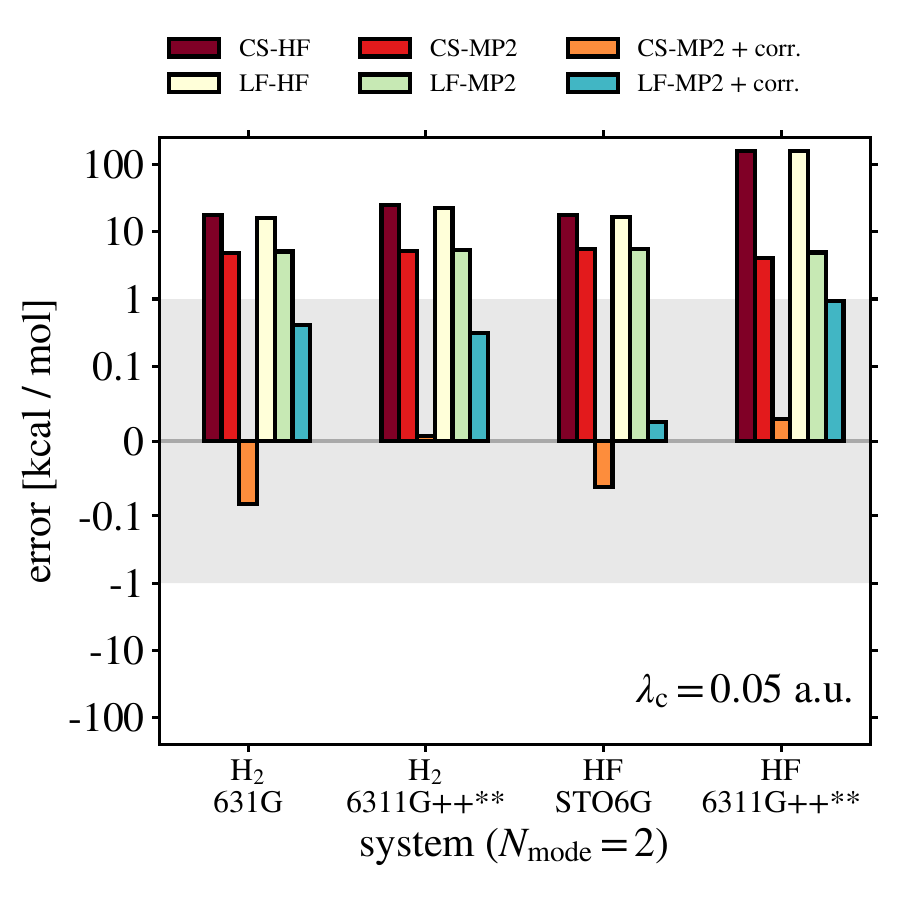}}
\caption{\label{fig: benchmark diatomic}Error in energy of diatomic molecules \ce{H2} and \ce{HF} in the medium-coupling regime ($\lambda_{\mathrm{c}} = 0.05$ a.u.). The reference energies of \ce{H2} are from exact diagonalization (ED) or DMRG (DMRG is used for the HF with 6311G++** basis). The shaded area labels chemical accuracy (error $< 1$ kcal/mol).}
\end{figure*}
Next, we benchmark the ground-state energy of diatomic molecules coupled to cavity photons with our methods. The Hamiltonian takes the form of Eqs.~\eqref{eq:ab initio g tensor}-\eqref{eq:ab initio h2}. To compare with the energies from ED/DMRG or QED-CCSD, we need to add a correction term from the missing energy of electron-electron interactions,
\begin{equation}\label{eq:correction elec}
\begin{split}
\Delta E^{\text{corr}} =  E^{\text{CCSD(T)\text{ or }CCSD}}[\Phi^{\elec}] - E^{\text{MP2}}[\Phi^{\elec}] ,
\end{split}
\end{equation}
where $\Phi^{\elec}$ is the ground-state electronic Slater determinant from CS-HF or LF-HF and this correction is added to the CS-MP2 or LF-MP2 energy.

 The photon frequencies are chosen to match the first electronic excited energy $\omega_0$ of the corresponding molecule,
\begin{equation}\label{eq:frequency multi mode}
\begin{split}
 \omega_x = (2x + 1) \omega_0 , \text{   } x = 1, 2, \cdots, N^{\text{mode}} ,
\end{split}
\end{equation}
where for hydrogen (\ce{H2}) $\omega^{\text{H}_2}_0 = 0.466751$ a.u. and for hydrogen fluoride (\ce{HF})  $\omega^{\text{HF}}_0 = 0.531916$ a.u.. The coupling constant $\lambda_{\text{c}}$ is chosen as 0.05 and 0.5 a.u. to assess medium and strong coupling situations respectively. \REV{We emphasize that the light-matter coupling used here is significantly larger than what can presently be achieved in experiments. For example, $\lambda_{\text{c}}$ = 0.1 corresponds to an effective mode volume of less than 0.2 nm$^3$ ($\lambda_{\text{c}} = \sqrt{\frac{1}{\epsilon_0 V}}$), which is less than the typical mode volumes $V$ of $\sim$30 nm$^3$ that can be achieved in plasmonic cavities~\cite{BenzS2016,MondalJPCL2022}. It is worth mentioning that there are exciting experimental developments of picocavities where effective mode volume are less than 1 nm$^3$ ~\cite{BenzS2016}. }

We first consider a single photon mode with an intermediate coupling $\lambda_{\text{c}} = 0.05$ a.u., as shown in Fig.~\ref{fig: benchmark diatomic} (a). The CS-HF and LF-HF energetics behave similarly for the \ce{H2} and \ce{HF} molecules.  In these cases, the MP2 corrections are also similar. After adding the electronic 
 correction  (Eq.~\eqref{eq:correction elec}), the errors from CS-MP2 and LF-MP2 are all within the chemical accuracy ($< 1$ kcal/mol). This reflects that these systems can be classified as weakly coupled problems (similar to the HH models with a small coupling constant), and are therefore very well described by CS-MP2 and LF-MP2.
 When more photon modes are considered (Fig.~\ref{fig: benchmark diatomic}(b)), the two methods are still similar and their errors are within the chemical accuracy.

\begin{figure}[!htb]
\includegraphics[width=0.48\textwidth, clip]{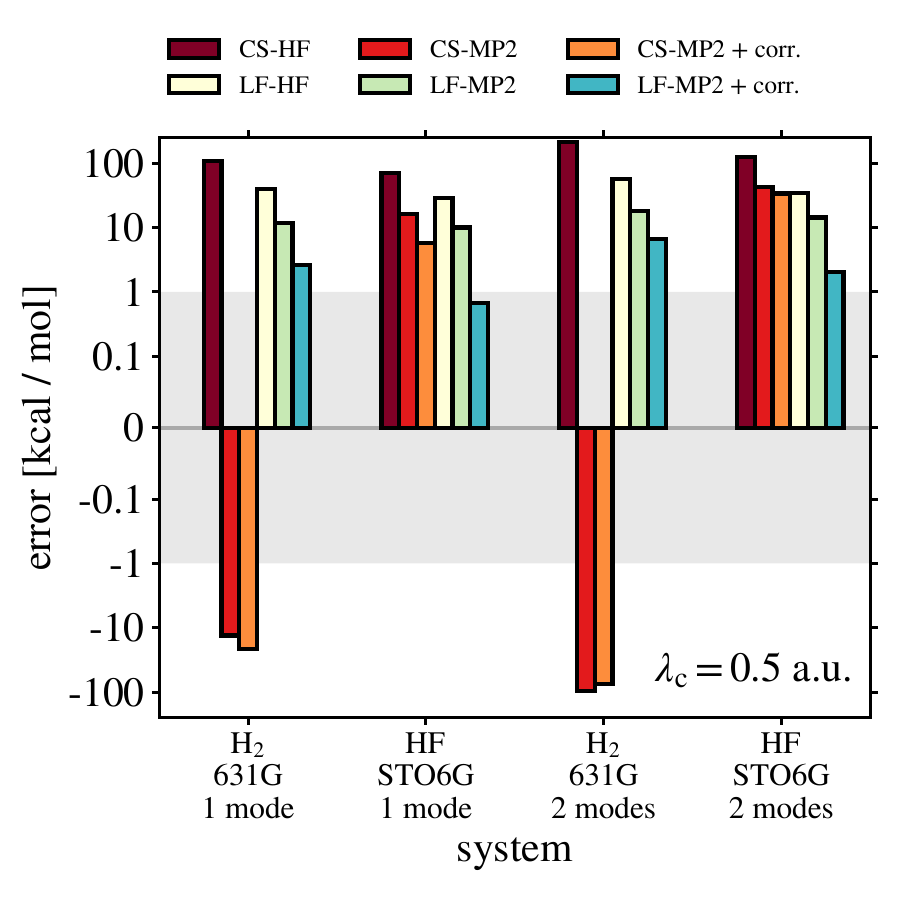}
\caption{\label{fig: benchmark lc 0.5}Error in the energy of diatomic molecules \ce{H2} and \ce{HF} in the strong-coupling regime ($\lambda_{\mathrm{c}} = 0.5$ a.u.). The reference energies are from exact diagonalization. The shaded area labels the chemical accuracy (error $< 1$ kcal/mol). }
\end{figure}
The large coupling case ($\lambda_{\text{c}} = 0.5$ a.u.), on the other hand, has larger errors for all methods. In particular, we see substantial improvements in LF-HF over CS-HF, and LF-MP2 over CS-MP2. For example, for \ce{H2} in 631G with 1 photon mode, the CS-MP2 with the correlation correction gives an error of $-21.3$ kcal/mol while 
LF-MP2 with the correction gives an error of $2.6$ kcal/mol.  This suggests that in the strong-coupling regime, the LF-type approaches are more accurate than the coherent-state-based ones. The difference between LF-HF and CS-HF can be regarded as a signal of strong coupling and large deviations indicate the degradation of the CS-based methods.

Furthermore, when more photon modes are included, the error also becomes larger, and in such cases, the LF-type approaches also provide larger improvements. In contrast to the intermediate coupling case, as the coupling becomes strong, the multi-mode effect becomes important and the single-mode approximation is no longer valid.  One needs to develop methods that explicitly include the higher photon modes with the awareness that such problems are potentially more strongly correlated.

\begin{figure*}[!htb]
\subfigure[]{\includegraphics[width=0.48\textwidth, clip]{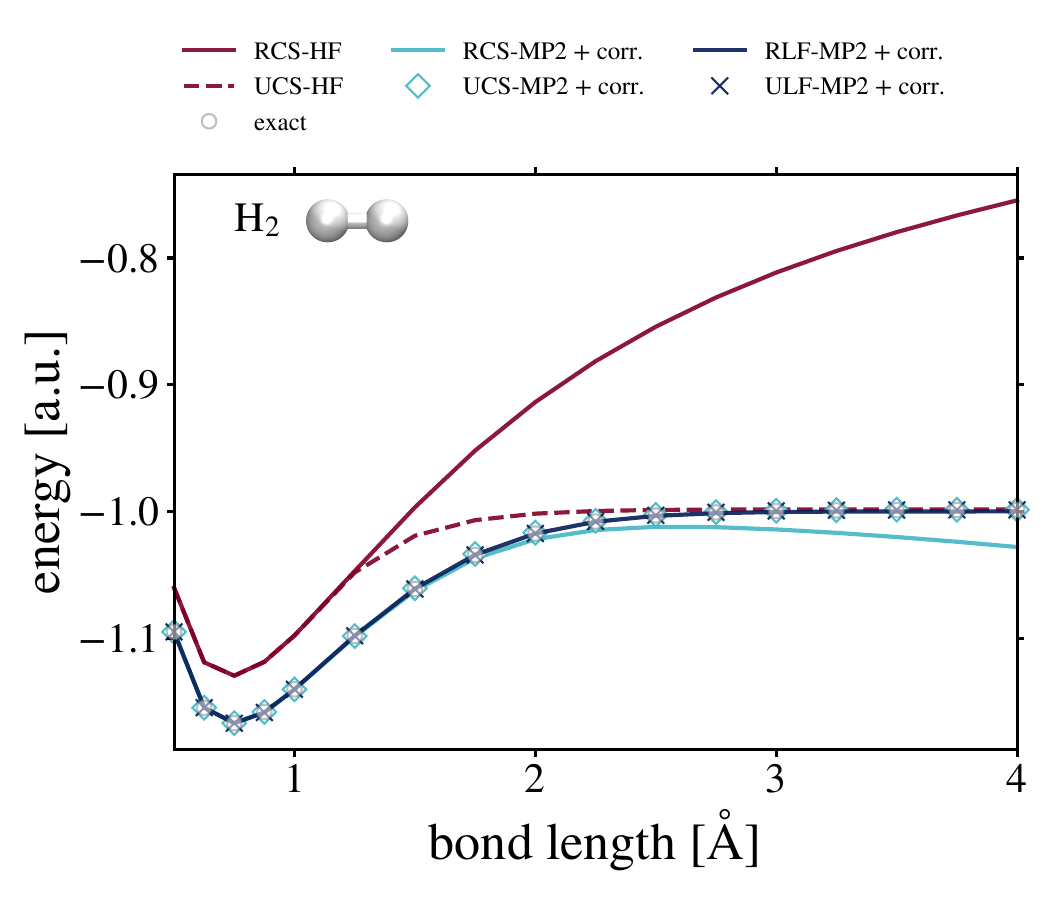}}
\subfigure[]{\includegraphics[width=0.48\textwidth, clip]{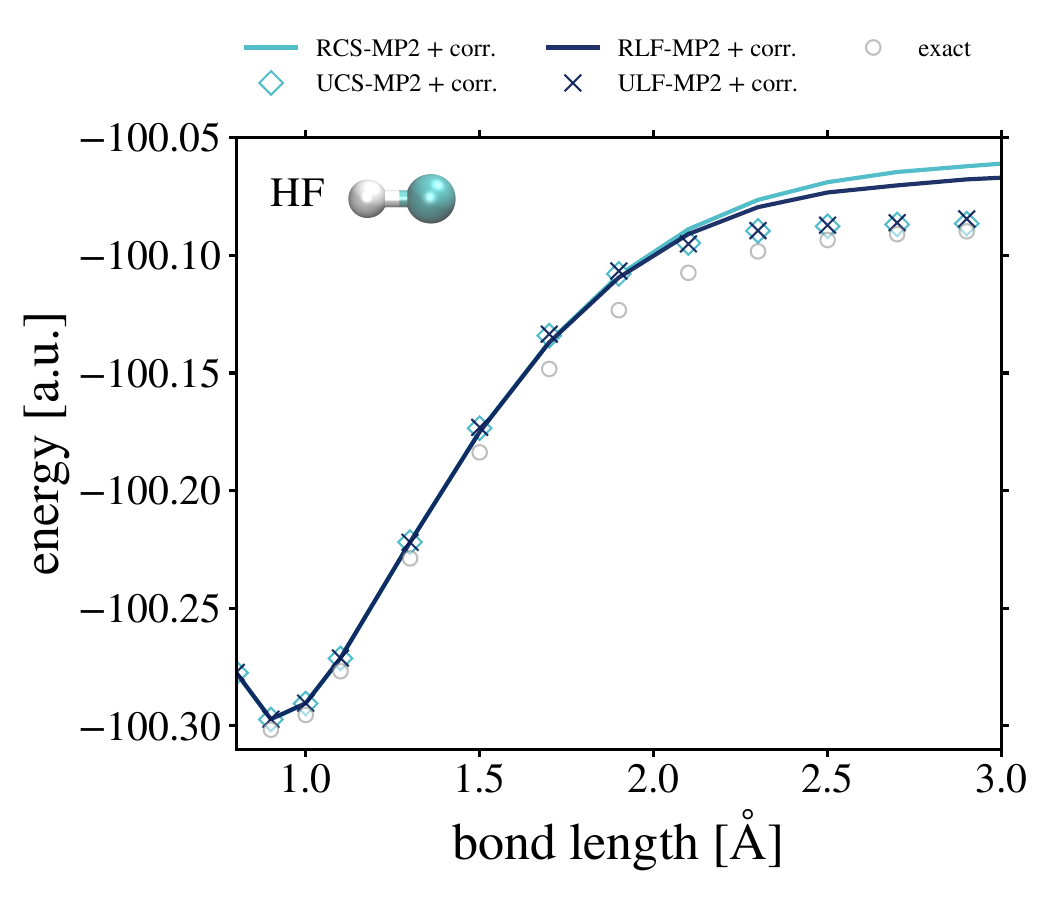}}
\caption{\label{fig: dissociation}Dissociation of diatomic molecules \ce{H2} and \ce{HF} in a photon cavity ($\lambda_{\mathrm{c}} = 0.05$ a.u.). Both spin-restricted (R) and unrestricted (U) determinants are considered.}
\end{figure*}
We finally discuss the dissociation of diatomic molecules in a photon cavity ($\lambda_{\mathrm{c}} = 0.05$ a.u.), as shown in Fig.~\ref{fig: dissociation}. In the \ce{H2} case, restricted (R) CS-HF performs similarly to regular HF, which does not capture the radical nature in the dissociation limit and thus exhibits a blow-up in energy. The unrestricted (U) HF does not break symmetry near the equilibrium geometry and overlaps with the RCS-HF. In the dissociation limit, the UCS-HF provide an accurate energy at the cost of breaking SU(2) $S^2$ symmetry, which is well-known in quantum chemistry. The perturbative correction on top of the unrestricted reference then gives an accurate description of the whole curve.  Interestingly, the RLF-MP2 + CCSD electronic correction (Eq.~\eqref{eq:correction elec}) gives almost exact results, whereas the RCS-MP2 + CCSD correction is lower in energy in the dissociation limit.  Since CCSD is exact for 2-electron systems, the error is likely from the electron-photon coupling part, and this reflects that the reference in the RCS-MP2 is not good enough for a low-order perturbative treatment of electron-photon interactions here.

In Fig.~\ref{fig: dissociation} (b), the dissociation of \ce{HF} molecule shows similar trends as those in \ce{H2} molecule.  Again, all curves overlap in the equilibrium region and deviate when the bond is stretched. The unrestricted references are expected to be good in the dissociation limit and hence give a flat result. The restricted reference, however, is not good for both electron-electron and electron-photon interactions in such a limit. The restricted curves are then shifted up a bit in energy, and the RLF-MP2 is slightly better than RCS-MP2.

\subsection{Proton-transfer reaction barriers}

\begin{figure*}[!htb]
\subfigure[]{\includegraphics[width=0.48\textwidth, clip]{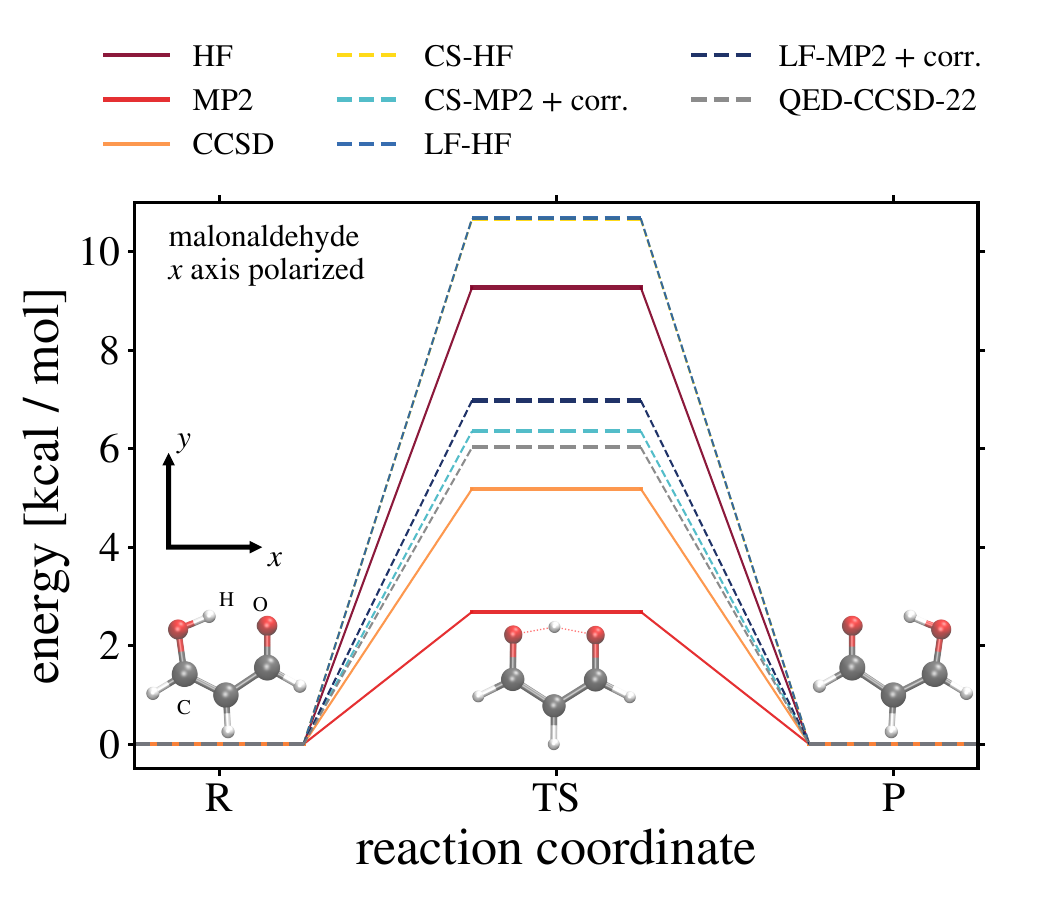}}
\subfigure[]{\includegraphics[width=0.48\textwidth, clip]{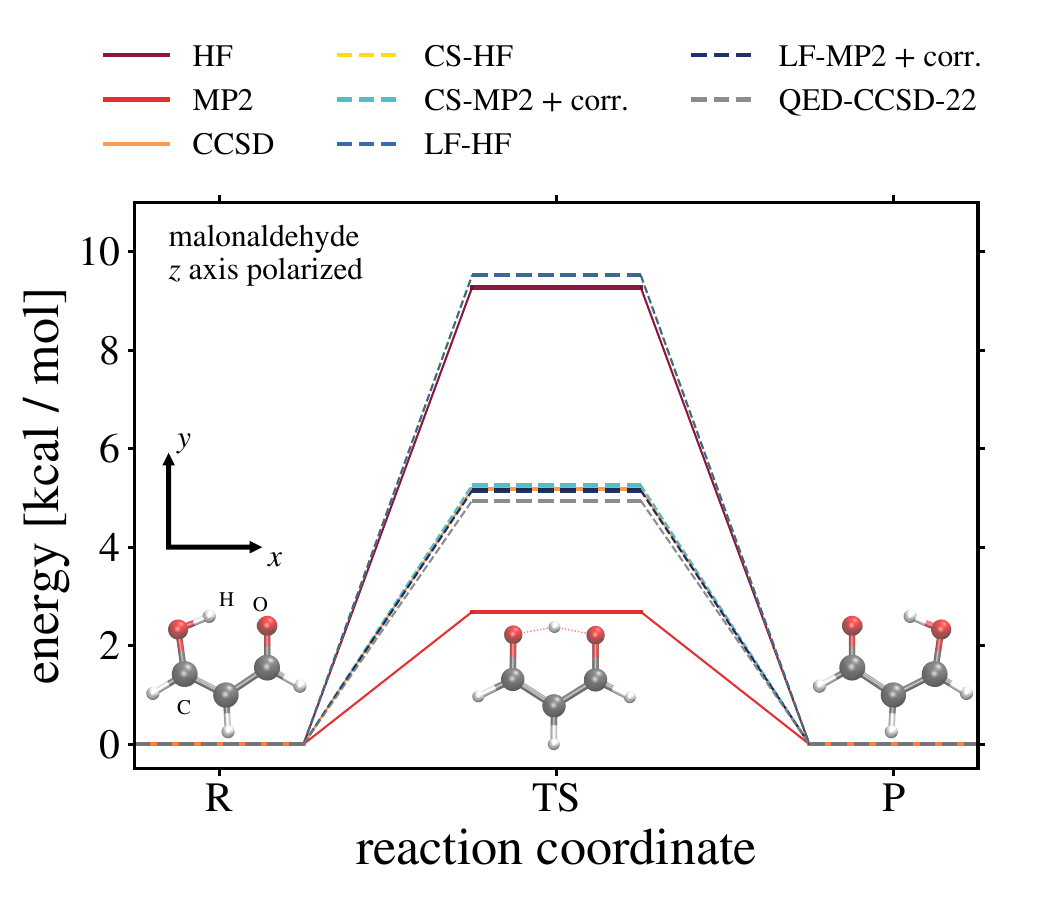}}
\caption{\label{fig:malon}Proton transfer in malonaldehyde inside and outside of a cavity. Only one photon mode is considered with $\omega_0 = 3$ eV and $\lambda_{\text{c}} = 0.1$ a.u. The HF, MP2, CCSD are methods without photon (and coupling) terms. QED-CCSD-22 is taken from Ref.~\cite{Pavosevic22-QED-CC-proton-transfer}. The polarization direction is along (a) the $x$ axis and (b) the $z$ axis. }
\end{figure*}
In this section, we consider two more realistic \abinitio systems and focus on the reaction barrier of proton-transfer reactions when they are coupled to optical cavities.

Malonaldehyde is a prototypical molecule in the study of proton transfer.  It is symmetric with respect to the reactant (R) and product (P). When the polarization direction is along the $x$ axis (Fig.~\ref{fig:malon} (a)), the existence of photons and their coupling to electrons increase the reaction barrier, which can be seen from the CS-HF/LF-HF versus the bare HF curve. The electron-electron correlation, on the other hand, reduces the reaction barrier. The MP2 and CCSD corrections lower the barrier by about 6 and 4 kcal/mol, respectively. In particular, the MP2 underestimates the barrier significantly and highlights the necessity of the correction term for the electron-electron interactions. Nevertheless, the electron-photon coupling only needs an MP2 level of theory. As we can see clearly from Fig.~\ref{fig:malon}(a), the CS-MP2 and LF-MP2 with electronic correction provide very similar barriers compared to the reference value from QED-CCSD-22~\cite{Pavosevic22-QED-CC-proton-transfer} (error $< 1$ kcal/mol). The similarity between CS and LF theories also suggests that the system should be within the weak-coupling region.
When the polarization direction is along the $z$ axis (Fig.~\ref{fig:malon} (b)), the electron-photon coupling shifts the barrier only slightly, and other properties are similar to the case along the $x$ axis.

\begin{figure*}[!htb]
\subfigure[]{\includegraphics[width=0.48\textwidth, clip]{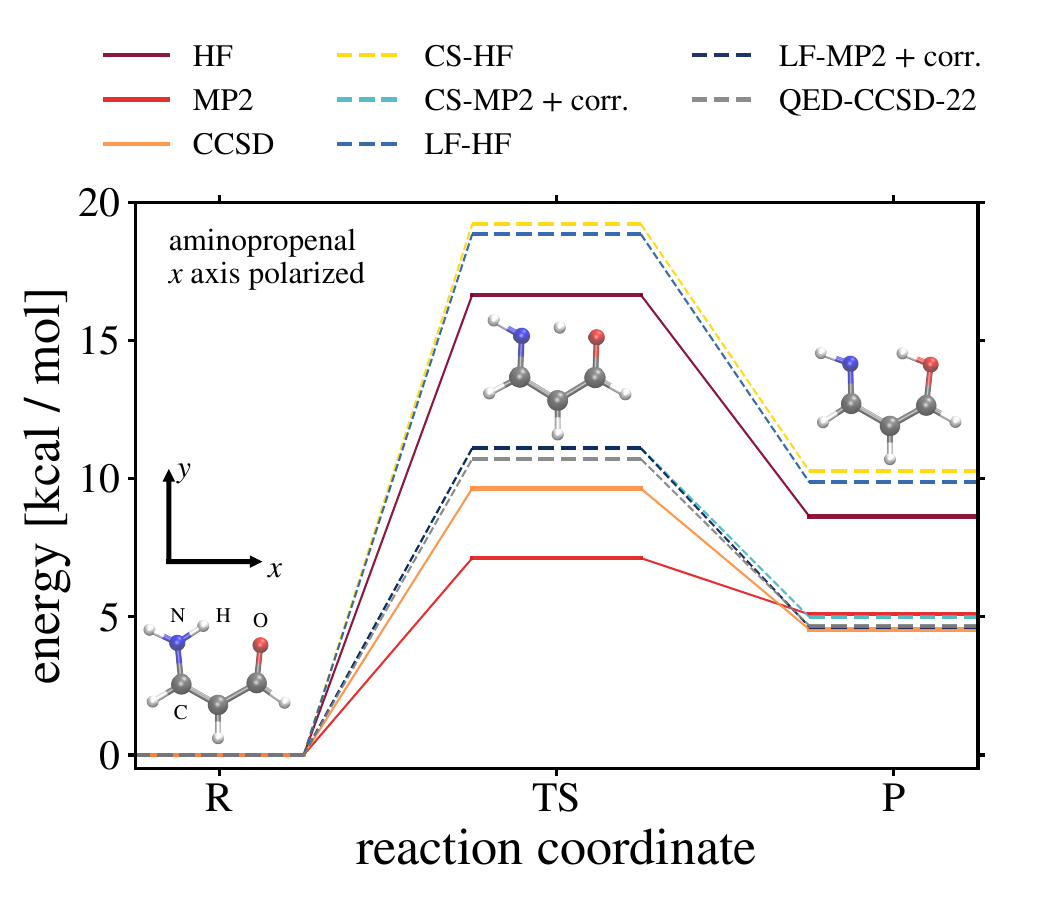}}
\subfigure[]{\includegraphics[width=0.48\textwidth, clip]{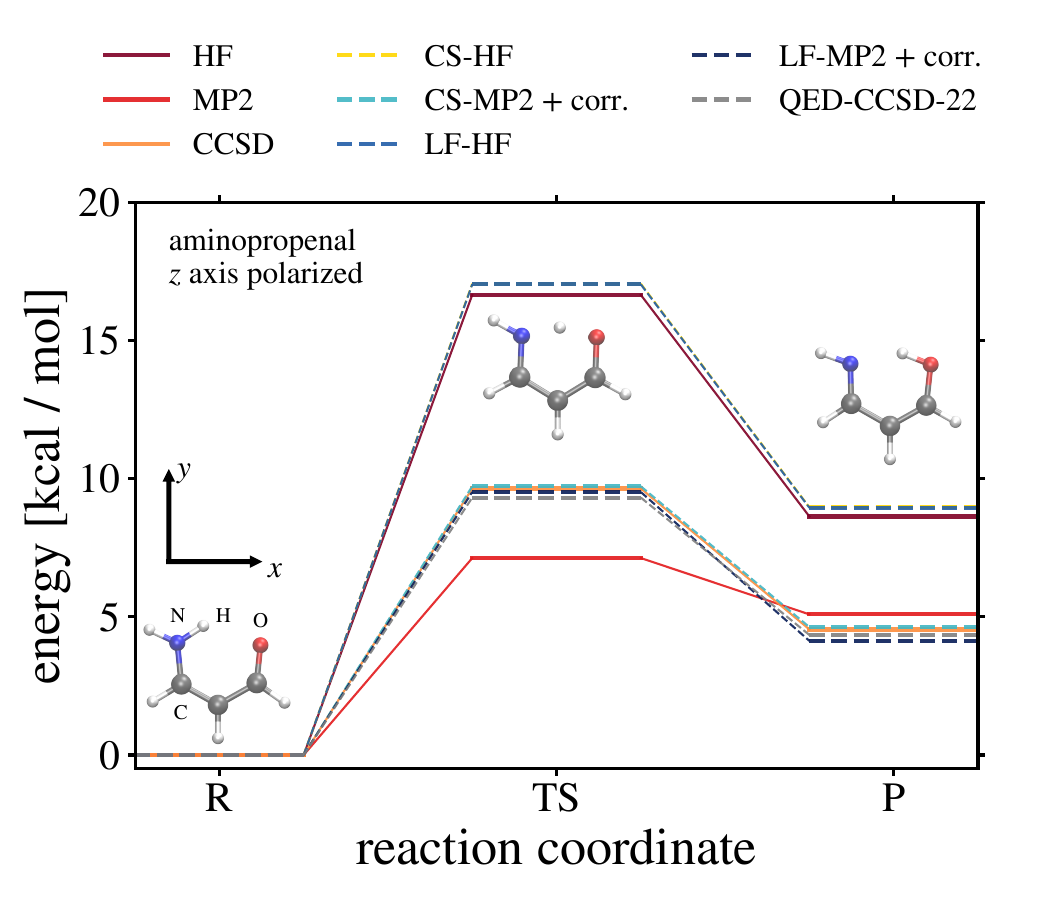}}
\caption{\label{fig: amino}Proton transfer in aminopropenal inside and outside of a cavity. Only one photon mode is considered with $\omega_0 = 3$ eV and $\lambda_{\text{c}} = 0.1$ a.u. The HF, MP2, and CCSD are methods without photon (and coupling) terms. QED-CCSD-22 is taken from Ref.~\cite{Pavosevic22-QED-CC-proton-transfer}. The polarization direction is along (a) the $x$ axis and (b) the $z$ axis.}
\end{figure*}
The Aminopropenal molecule is asymmetric with respect to reactants and products. The results are collected in Fig.~\ref{fig: amino}. For the reaction barrier, the observations are similar to those observed with malonaldehyde. Here, CS-HF and LF-HF also overestimate the reaction energy ($E_{\text{P}} - E_{\text{R}}$), which is corrected by correlation terms.
The LF-based methods are slightly better in the description of the reaction energy than the CS-based methods.

\section{Conclusions}\label{sec: conclusion}

In summary, we developed a Lang-Firsov transformation-based perturbation scheme for electron-boson coupled problems. The method relies on the variational optimization of Lang-Firsov and coherent parameters for a zero-boson averaged electronic Hamiltonian, and the subsequent use of a M\o{}ller-Plesset perturbative treatment for electron-boson couplings.

As shown in the cases of the Hubbard-Holstein model and \abinitio molecules (diatomic molecules and proton transfer reaction barriers coupled to a photon cavity), the LF-based methods perform well for weakly-coupled systems and strongly-coupled limits. The methods provide a natural interpolation between the two limits. 
The comparison between CS and LF-based approaches is useful for identifying the transition to the strong electron-boson couplings regime, and in these systems, the LF-based methods are preferred.
When the coupling constant becomes extremely large or when multi-boson modes are involved, the CS-based methods should be used with caution, and a more generalized Lang-Firsov transformation may be useful for capturing the non-diagonal effects.

In the intermediate coupling region, the LF-MP scheme may underestimate the correlation energy due to the symmetry-breaking from the variational optimization. This is in principle a multi-reference problem, and a CI-based or multi-reference-based perturbation theory should be used. Another possibility is to introduce a symmetry-breaking electronic reference. As shown by the dissociation curve of the diatomic molecules, a spin-symmetry breaking reference improves the overall energy description. Other types of symmetry-breaking schemes are possible and may be useful for condensed matter systems (such as a Bardeen–Cooper–Schrieffer reference for superconductors coupled to phonons or photons). We plan to explore these and related directions in future works.

\begin{acknowledgments}
We thank Garnet Chan, Ankit Mahajan, Junjie Yang, Huanchen Zhai, and Chong Sun for the helpful discussions. ZHC was funded by the Columbia MRSEC on Precision-Assembled Quantum Materials (PAQM) under award number DMR-2011738.

\REV{Data availability. The data supporting this study's findings are presented in the manuscript or available from the corresponding author upon request.}

\REV{Code availability. The developed code used for this
study is publicly available on Github~\cite{polar-code}.}


\end{acknowledgments}

\appendix

\section{Uniform and generalized Lang-Firsov transformation}\label{app: different form of LF}

Here, we discuss other possible forms of Lang-Firsov transformation. 

\begin{figure*}[!htb]
\subfigure[]{\includegraphics[width=0.48\textwidth, clip]{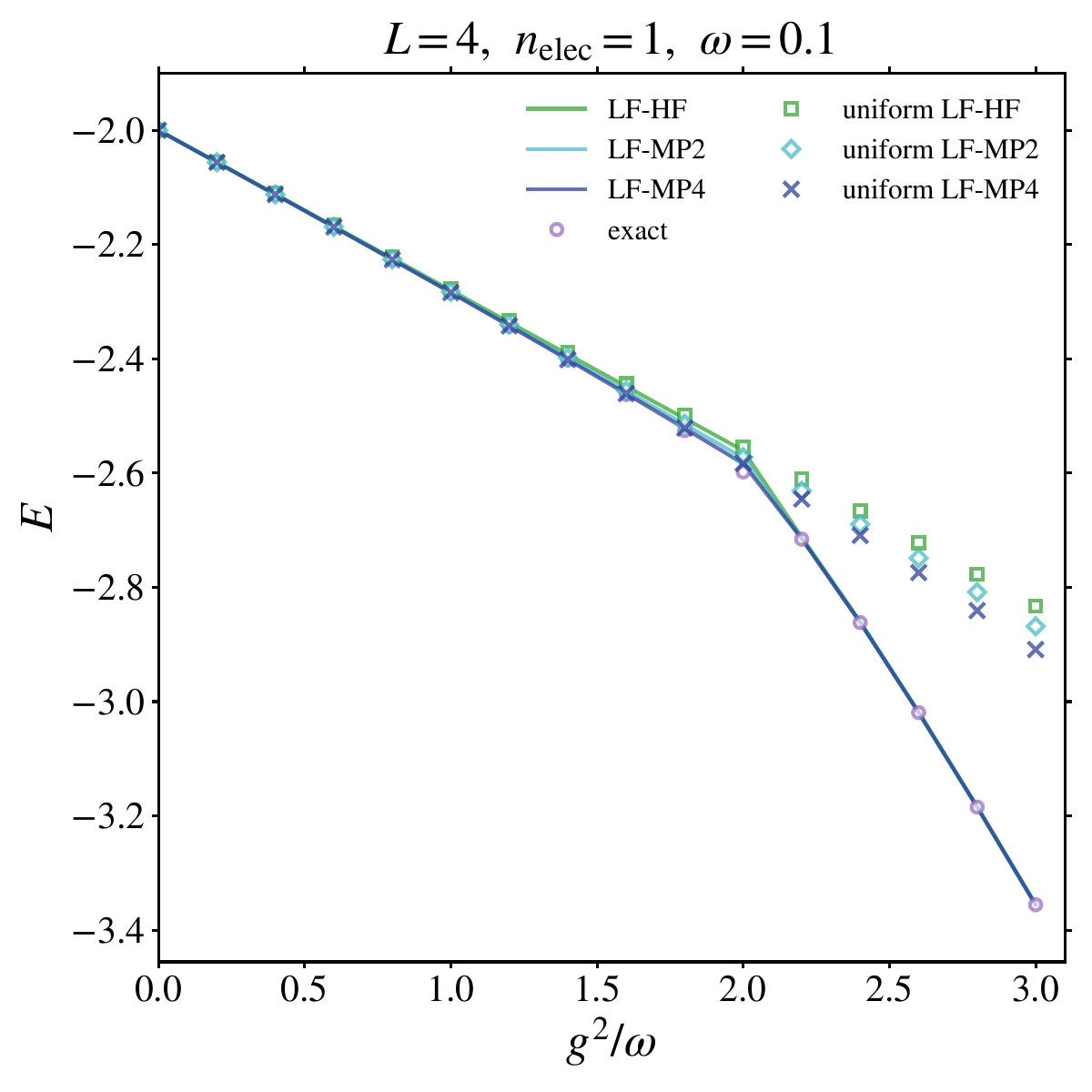}}
\subfigure[]{\includegraphics[width=0.48\textwidth, clip]{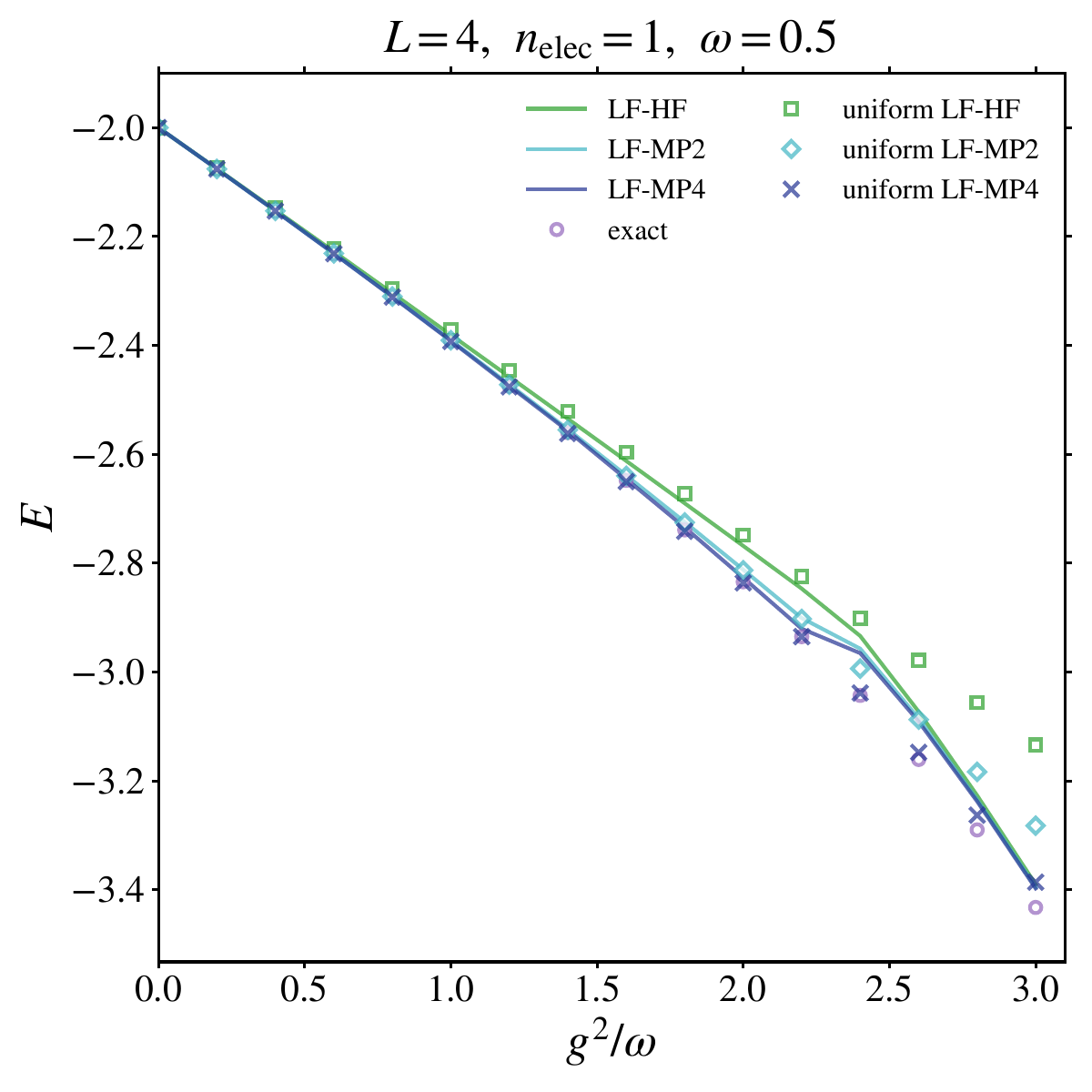}}
\caption{\label{fig: hh model uniform} The uniform version of LF transform is compared to the non-uniform one ($\lambda$ vs. $\lambda^x_p$), with two different model frequencies (a) $\omega = 0.1$ and (b) $\omega = 0.5$. }
\end{figure*}
The simplest version is the uniform version (the original parametrization in the Ref.~\cite{Lang63-Lang-Firsov-trans}), where every site in the Hubbard-Holstein model shares the same scalar value of $\lambda$ and $z$, which we call {\textit{uniform LF}}. The results are shown in Fig.~\ref{fig: hh model uniform}. It can be seen that in the weak coupling regime, there is no symmetry breaking, and the uniform LF and LF are almost identical. In the $\omega = 0.1$ case,  the uniform MP follows roughly a straight line, which does not capture the slope change. While in the $\omega = 0.5$ case, the uniform LF-MP behaves better near the transition point around $g^2/\omega = 2.2$. This result reflects that in the strongly coupling region, a perturbation starting from a symmetry-preserved state may give larger corrections, and in some cases (such as the intermediate coupling region at $\omega = 0.5$), the uniform LF-MP gives better results.

A more {\textit{general Lang-Firsov (GLF)}} transformation, which includes coupling to a non-diagonal electronic density, is
\begin{equation}\label{eq: GLF transform U}
\hat{U}^{\GLF} = \exp[\sum_{xpq} \lambda^{x}_{pq} a^{\dg}_{p} a_{q} (b_x - b^{\dg}_x)] .
\end{equation}
One can use the BCH expansion to obtain the  effect of this transformation on boson operators,
\begin{equation}
\hat{U}^{\GLF \dg} b_{x} \hat{U}^{\GLF} = b_x - \sum_{pq} \lambda^{x}_{pq} a^{\dg}_{p} a_{q} .
\end{equation}
The transformation of the electron operator, on the other hand, involves the combination of all other operators, 
\begin{equation}
\begin{split}
\hat{U}^{\GLF \dg} a_{q} \hat{U}^{\GLF} 
&= \sum_{p} \qty[\ee^{\sum_{y}\lambda^{y}_{pq} (b_y - b^{\dg}_y)}]_{qp} a_p  .
\end{split}
\end{equation}
The transformed Hamiltonian after zero-boson averaging can be expressed in the BCH expansion (up to $2N$-th order),
\begin{equation}
\begin{split}
\hat{H}^{\GLF, \elec} =& \mel{0^{\text{p}}}{\hat{U}^{\GLF \dg}\hat{H} \hat{U}^{\GLF} }{0^{\text{p}}} .\\
\end{split}
\end{equation}
The Hamiltonian elements are
\begin{equation}
\begin{split}\label{eq:GLF H exact}
H^{\GLF} = & \sum^{2N}_{n=2} (-1)^{N} \frac{n_{x_1} !! \cdots n_{x_n} !!}{n!} \sum^{\text{all strings of order $n$}}_{{x_1, x_2, ..., x_n}} \\
&\times \qty[[\cdots \qty[H, \lambda^{x_1}], \cdots, \lambda^{x_{n-1}}] \lambda^{x_n}] ,
\end{split}
\end{equation}
where the one-body and two-body commutators are defined as,
\begin{equation}
\begin{split}
[h, \lambda^{x}]_{pq} = \qty(h \cdot \lambda^{x} - \lambda^{x} \cdot  h)_{pq} ,
\end{split}
\end{equation}
\begin{equation}
\begin{split}
[V, \lambda^{x}]_{pqrs} = \sum_{m} & \left(V_{pmrs} \lambda^{x}_{mq} + V_{pqrm} \lambda^{x}_{ms} \right. \\
 &- \left. V_{mqrs} \lambda^{x}_{mp} - V_{pqms} \lambda^{x}_{mr} \right) .
\end{split}
\end{equation}
The BCH expansion in Eq.~\eqref{eq:GLF H exact}, however, only converges when $\lambda$ is small, and the number of boson strings scales exponentially. Thus, it is not practical to apply.

One approximation proposed in Refs.~\cite{Hannewald04-canonical-transform, Luo22-canonical-transform} is to assume $\lambda$'s of different mode $x$ commute with each other, and this leads to a simplification of the nested commutator and can be evaluated to the infinite order such that the restriction on small $\lambda$ is relaxed.
\begin{equation}
\begin{split}
h^{\GLF}_{pq} = \sum_{ij} \qty(\ee^{\Lambda})_{pq, ij} h_{ij} , \\
\end{split}
\end{equation}
\begin{equation}\label{eq:GLF exp h2}
\begin{split}
V^{\GLF}_{pqrs} = \sum_{ijkl} \qty(\ee^{\Lambda})_{pqrs, ijkl} V_{ijkl} , \\
\end{split}
\end{equation}
where $\Lambda$ is defined as the double commutator operator, whose action on a Hamiltonian $H$ is defined as
\begin{equation}\label{eq:double commutator}
\begin{split}
\Lambda H = -\frac{1}{2}\qty[\qty[H, \lambda^{x}], \lambda^{x}] .
\end{split}
\end{equation}
Note the Eq.~\eqref{eq:GLF exp h2} formally has computational complexity $\mathcal{O}(N^{8}_{\text{orb}})$ as well as a impractical storage of a very large $\mathcal{O}(N^{8}_{\text{orb}})$ exponential tensor. This bottleneck can be circumvented by the algorithm of computing the action of a matrix exponential on a vector~\cite{Al11-expm-multiply}, which does not require the explicit build of the  $\mathcal{O}(N^{8}_{\text{orb}})$ tensor. By using such an algorithm, only several evaluations of the double commutator (Eq.~\eqref{eq:double commutator}) are needed, and thus scales only $\mathcal{O}(N^{5}_{\text{orb}})$. Furthermore, the storage is only $\mathcal{O}(N^{4}_{\text{orb}})$.

\begin{table}[!htb]
\caption{\label{tab:compare lf and glf}
Comparison of uniform LF, LF, and generalized LF (GLF) transformation. The systems include the Hubbard-Holstein (HH) models ($\omega = 0.5$) with one electron, and the \ce{H2} molecule (in 631G basis with a different coupling constant $\lambda_{\text{c}}$). The energies/coupling constant for \abinitio systems are in the unit of a.u. The uniform LF-HF is meaningless for \abinitio systems since the orbitals are not translational invariant.}
\begin{ruledtabular}
\begin{tabular}{lcccccccc}
 System & CS-HF & LF-HF & LF-HF & GLF-HF \\
        &       & uniform &  & \\
\hline
HH model ($\frac{g^2}{\omega} = 1.0$) & -2.2500 & -2.3711 & -2.3807 & -2.3808  \\
HH model ($\frac{g^2}{\omega} = 2.4$) & -2.6000 & -2.9016 & -2.9339 & -2.9435  \\
\ce{H2} (1 mode, $\lambda_{\text{c}} = 0.05$) &  -1.1239 & - & -1.1252 & -1.1253 \\
\ce{H2} (1 mode, $\lambda_{\text{c}} = 0.5$) &  -0.8710 & - & -0.9790 & -0.9904 \\
\ce{H2} (2 modes, $\lambda_{\text{c}} = 0.05$) &  -1.1212 & - & -1.1245 & -1.1246 \\
\ce{H2} (2 modes, $\lambda_{\text{c}} = 0.5$) & -0.6432   & - & -0.8991  & -0.9244 \\
\end{tabular}
\end{ruledtabular}
\end{table}
In Table~\ref{tab:compare lf and glf}, we compare the performance of uniform LF, LF, and GLF on the Hubbard-Holstein model and \ce{H2} molecule in a cavity. 

In the Holstein model, because of the special local structure of the electron-phonon coupling tensor (Eq.~\eqref{eq:hh model g tensor}), the improvement from the non-local coupling (GLF) is marginal (the correction is slightly larger for the stronger coupling case, $g^2/\omega = 2.4$).

 For the \ce{H2} molecule with a small coupling constant $\lambda_{\text{c}} = 0.05$, the energy from GLF-HF is only slightly lower than that from LF-HF. The two solutions are almost identical. For the large coupling $\lambda_{\text{c}} = 0.5$, however, the correction from GLF is larger. As discussed in the main text, the multi-mode effect increases the correlation energy of electron-photon coupling. Such a strongly coupled system naturally has larger correction from a larger variational space.
In general, for \abinitio systems, the electron-boson coupling tensor is no longer diagonal. Although in this work LF-MP2  (with an electronic correlation correction) appears to already work well for most systems, a generalized transformation is expected to be useful for 
 extremely strongly coupled systems.

\section{Analytical gradient of variational Lang-Firsov approach}\label{app:analytical grad}

By applying the chain rule to the LF-HF energy (Eq.~\eqref{eq:LFHF total energy}), one can compute its gradient with respect to variational parameters.

The  gradient of coherent shifts $z_x$ is,
\begin{equation}
\begin{split}
&\pdv{E}{z^{y}} = 2 \omega_y z_y + 2 \sum_{pq} g^{y}_{pq} e^{\Lambda_{pq}} \gamma_{qp} - 2 \sum_{p} \omega_y \lambda^{y}_{p} \gamma_{pp} ,
\end{split}
\end{equation}
where we defined the shorthand notations,
\begin{equation}
\begin{split}
\Lambda_{pq} &= -\frac{1}{2} \sum_{y}(\lambda^{y}_{q} - \lambda^{y}_{p})^2 ,\\
\end{split}
\end{equation}
\begin{equation}
\begin{split}
\Lambda_{pqrs} &= -\frac{1}{2} \sum_{y}(\lambda^{y}_{q} + \lambda^{y}_{s} - \lambda^{y}_{p} - \lambda^{y}_{r})^2 .
\end{split}
\end{equation}
The (spin-restricted) gradient of LF parameters $\lambda^{y}_{m}$ are (Einstein summation is assumed for clarity),
\begin{equation}
\begin{split}
	&\pdv{E}{\lambda^{y}_{m}}  =
	2 \lambda^{y}_{p}  h^{\eff}_{pq} \ee^{\Lambda_{pq}} \gamma_{qm}  -2 \lambda^{y}_{m}  h^{\eff}_{pq} \ee^{\Lambda_{pq}} \gamma_{qm} \\
	& -2 g^{y}_{pq} \ee^{\Lambda_{pq}} \gamma_{qp}  +2 \omega_y  \lambda^{y}_{m} z_y \gamma_{mm} \\
	& + 2 \omega_y \lambda^{y}_{q} \gamma_{qq} \gamma_{mm}  - \omega_y \lambda^{y}_{q} \gamma_{mq} \gamma_{mq}  \\
	& - 2 V_{pmrs} \ee^{\Lambda_{pmrs}} \lambda^{y}_{m} \gamma_{mp} \gamma_{sr}  + V_{pmrs} \ee^{\Lambda_{pmrs}} \lambda^{y}_{m} \gamma_{ps} \gamma_{mr} \\
	& + 2 V_{pmrs} \ee^{\Lambda_{pmrs}} \lambda^{y}_{p} \gamma_{mp} \gamma_{sr}  - V_{pmrs} \ee^{\Lambda_{pmrs}} \lambda^{y}_{p} \gamma_{ps} \gamma_{mr} \\
	& - 2 V_{pmrs} \ee^{\Lambda_{pmrs}} \lambda^{y}_{s} \gamma_{mp} \gamma_{sr}  + V_{pmrs} \ee^{\Lambda_{pmrs}} \lambda^{y}_{s} \gamma_{ps} \gamma_{mr} \\
	& + 2 V_{pmrs} \ee^{\Lambda_{pmrs}} \lambda^{y}_{r} \gamma_{mp} \gamma_{sr}  - V_{pmrs} \ee^{\Lambda_{pmrs}} \lambda^{y}_{r} \gamma_{ps} \gamma_{mr} \\
	& -4 g^{y}_{pm} \ee^{\Lambda_{pm}} \lambda^{y}_{r}  \lambda^{y}_{m} \gamma_{mp} \gamma_{rr}  +2 g^{y}_{pm} \ee^{\Lambda_{pm}} \lambda^{y}_{r}  \lambda^{y}_{m} \gamma_{pr} \gamma_{mr} \\
	& +4 g^{y}_{mq} \ee^{\Lambda_{mq}} \lambda^{y}_{r}  \lambda^{y}_{q} \gamma_{qm} \gamma_{rr}  -2 g^{y}_{mq} \ee^{\Lambda_{mq}} \lambda^{y}_{r}  \lambda^{y}_{q} \gamma_{mr} \gamma_{qr} \\
	& - 2 g^{y}_{pq} \ee^{\Lambda_{pq}} \gamma_{qp} \gamma_{mm}  +   g^{y}_{pq} \ee^{\Lambda_{pq}} \gamma_{pm} \gamma_{qm} . 
\end{split}
\end{equation}

The gradient of orbital rotation $\kappa_{ia}$ is the occupied-virtual block of the Fock matrix, 
\begin{equation}
\begin{split}
&\pdv{E}{\kappa_{ia}} = F_{ia} .
\end{split}
\end{equation}

\section{Different choice of dipole self-energy}\label{app:dipole self energy}

\begin{table}[H]
\caption{\label{tab:dipole self-energy}
The influence of different forms of dipole self-energy (product of second-quantized operator Eq.~\eqref{eq:ab initio h1} vs second-quantization after product Eq.~\eqref{eq:ab initio h1 quadrupole}). The basis setting is the same as in Fig.~\ref{fig: benchmark diatomic}.}
\begin{ruledtabular}
\begin{tabular}{lcccccccc}
 System &  \multicolumn{2}{c}{Eq.~\eqref{eq:ab initio h1}} &  \multicolumn{2}{c}{Eq.~\eqref{eq:ab initio h1 quadrupole}}\\
        &    LF-HF   & LF-MP2   & LF-HF & LF-MP2 \\
\hline
\ce{H2} ($\lambda_{\text{c}} = 0.05$, 631G) & -1.12525 & -1.14263 & -1.12522 & -1.14259 \\
\ce{H2} ($\lambda_{\text{c}} = 0.05$, 6311G++**)  & -1.13105 & -1.15918 &  -1.13105 & -1.15917\\
\ce{H2} ($\lambda_{\text{c}} = 0.5$, 631G) & -0.97902 & -1.02336 & -0.96022 & -1.00246 \\
\end{tabular}
\end{ruledtabular}
\end{table}
\REV{In Table~\ref{tab:dipole self-energy}, we compare the different choices of dipole self-energy. The energy difference is more evident when small basis/large coupling constant/correlation methods are applied.}


\bibliography{lang_firsov_mp}

\end{document}